\documentclass[prb,aps,showpacs,twocolumn,amsmath,amssymb,floatfix,superscriptaddress]{revtex4-2}

\usepackage[unicode=true, colorlinks=true, citecolor={blue!80!black}, urlcolor={blue!50!black}, linkcolor = {blue!80!black}]{hyperref}
\usepackage{amsmath}
\usepackage{amssymb}
\usepackage{graphicx}
\usepackage{bm}
\usepackage{dsfont}
\usepackage{xcolor}
\usepackage[utf8]{inputenc}
\usepackage[normalem]{ulem}
\usepackage{physics}

\newcommand{\beq}{\begin{equation}}
\newcommand{\eeq}{\end{equation}}
\newcommand{\beqa}{\begin{eqnarray}}
\newcommand{\eeqa}{\end{eqnarray}}

\begin{document}

\title{Fluxoid valve effect in full-shell nanowire Josephson junctions}

\author{Carlos Payá}
\affiliation{Instituto de Ciencia de Materiales de Madrid (ICMM), CSIC, Madrid, Spain}
\author{F.J. Matute-Cañadas}
\affiliation{Departamento de Física Teórica de la Materia Condensada, Condensed Matter Physics Center (IFIMAC) and
Instituto Nicolás Cabrera, Universidad Autónoma de Madrid, Madrid, Spain}
\author{A. Levy Yeyati}
\affiliation{Departamento de Física Teórica de la Materia Condensada, Condensed Matter Physics Center (IFIMAC) and
Instituto Nicolás Cabrera, Universidad Autónoma de Madrid, Madrid, Spain}
\author{Ramón Aguado}
\affiliation{Instituto de Ciencia de Materiales de Madrid (ICMM), CSIC, Madrid, Spain}
\author{Pablo San-Jose}
\affiliation{Instituto de Ciencia de Materiales de Madrid (ICMM), CSIC, Madrid, Spain}
\author{Elsa Prada}
\email{elsa.prada@csic.es}
\affiliation{Instituto de Ciencia de Materiales de Madrid (ICMM), CSIC, Madrid, Spain}

\date{\today}

\begin{abstract}
We introduce a new type of supercurrent valve based on full-shell nanowires. These hybrid wires consist of a semiconductor core fully wrapped in a thin superconductor shell and subjected to an axial magnetic field. Due to the tubular shape of the shell, the superconductor phase acquires an integer number $n$ of $2\pi$ twists or \textit{fluxoids} that increases in steps with applied flux. By connecting two such hybrid wires, forming a Josephson junction (JJ), a flux-modulated supercurrent develops. If the two superconducting sections of the JJ have different radii $R_1$ and $R_2$, they can develop equal or different fluxoid numbers $n_1,n_2$ depending on the field. If $n_1\neq n_2$ the supercurrent is blocked, while it remains finite for $n_1=n_2$. This gives rise to a fluxoid valve effect controlled by the applied magnetic field or a gate voltage at the junction. We define a fluxoid-valve quality factor that is perfect for cylindrically symmetric systems and decreases as this symmetry is reduced. We further discuss the role of Majorana zero modes at the junction when the full-shell nanowires are in the topological superconducting regime.
\end{abstract}

\maketitle

\section{Introduction}

In nanoelectronics, a valve is a device that regulates (opens or closes) the flow of an electrical current through a nanometric circuit by changing a control parameter. It serves to encode and manipulate information stored in the electronic charge~\cite{Haselman:PI10}. Over the last decades, there has been significant progress in the manipulation of other electron degrees of freedom, such as the spin. The prototypical device in this case is the spin valve \cite{Parkin:PI03,Zutic:RMP04}, which in its simplest form consists of a non-magnetic material sandwiched between two ferromagnets. In this system, the propagation of a spin-polarized current depends on the relative magnetization orientation of the two magnetic layers, which can be controlled e.g. by an external magnetic field. There are other valves based on the electron spin, such as the superconducting spin-valve \cite{Niedzielski:PRB18,Stoddart-Stones:CP22,Bobkov:PRB24} or the magnon valve \cite{Wu:PRL18}. All of these have important applications in spintronics \cite{Zutic:RMP04}. More exotic types of valves are based on other quantum degrees of freedom. These include pseudospin \cite{San-Jose:PRL09,Jung:NM20} or valley \cite{Li:S18} valves in two-dimensional crystals, foundational for the fields of pseudospintronics \cite{Pesin:NM12,Bao:NM20} and valleytronics \cite{Schaibley:NRM16}, or all-optical pseudospin valves in nonlinear photonic crystals \cite{Izhak:OLO24}, useful for quantum optical information processing. 

In this work, we demonstrate a new type of superconducting \emph{fluxoid} valve effect  that is possible in Josephson junctions (JJs) based on full shell nanowires \cite{Goffman:NJP17, Tosi:PRX19, Matute-Canadas:PRL22, Giavaras:PRB24,Paya:PRB25}. These hybrid nanowires have come to the spotlight in recent years, both from a theoretical \cite{Woods:PRB19, Penaranda:PRR20, Kopasov:PSS20, Kopasov:PRB20, Metzger:PRR21, Escribano:PRB22, San-Jose:PRB23, Paya:PRB24, Paya:PRB24a,Vezzosi:SP25} and experimental \cite{Vaitiekenas:PRB20, Vaitiekenas:S20, *Vaitiekenas:ErrS25, Valentini:S21, Valentini:N22, Valentini:PRR25, Deng:PRL25} point of view, because of their relevance in the context of the Little-Parks (LP) effect~\cite{Little:PRL62,Parks:PR64,Tinkham:96} and topological superconductivity~\cite{Qi:RMP11,Alicea:RPP12, Aguado:RNC17, Vaitiekenas:PRB20, Vaitiekenas:S20, *Vaitiekenas:ErrS25, Prada:NRP20, Vekris:SR21, Ibabe:NC23, Ibabe:NL24}. They consist of a semiconductor core fully wrapped by a thin superconductor shell and are subjected to an axial magnetic field $B$. They typically have a hexagonal cross section~\cite{Ercolani:N12, Krogstrup:NM15, Vaitiekenas:PRB20}, but it has been shown that assuming cylindrical symmetry is a good approximation for low-density semiconductor cores at low energies~\cite{Paya:PRB24}. In this case, the applied flux is $\Phi=\pi R^2 B$, where $R$ is the shell mid radius. Due to the doubly connected geometry of the shell and the single-valuedness of the superconductor order parameter, its phase acquires, at equilibrium, a quantized winding number $n$ around the shell axis with flux. The number $n\in \mathbb{Z}$ of $2\pi$ twists is called the fluxoid number~\cite{Little:PRL62, Parks:PR64}, and it grows in steps with flux as $n(\Phi) = \lfloor \Phi /\Phi_0 \rceil$, where $\Phi_0=h/2e$ is the superconducting flux quantum and $\lfloor \rceil$ represents a rounding to the closest integer.

\begin{figure}
   \centering
   \includegraphics[width=0.9\columnwidth]{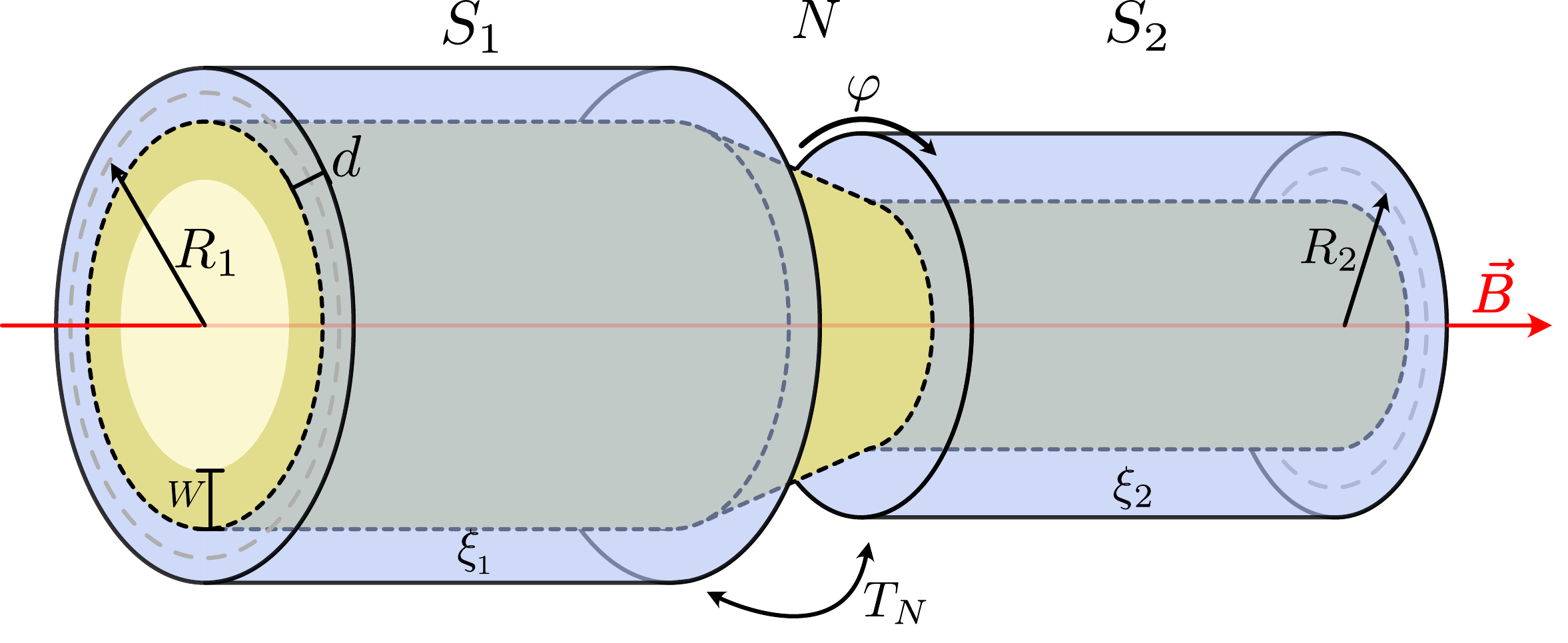}
   \caption{Sketch of a S$_1$NS$_2$ short JJ between two full-shell hybrid nanowires in a cylindrical approximation with different mid radii $R_1$ and $R_2$. The semiconductor core is represented in yellow and the thin superconductor shells of thickness $d$ in blue. Radially, most of the core charge density is concentrated close to the superconductor-semiconductor interface in an accumulation layer of thickness $W$. S$_1$ and S$_2$ may have different diffusive coherence lengths $\xi_1$ and $\xi_2$. An axial magnetic field $B$ is applied. $T_{\rm{N}}$ is the normal transmission that characterizes the weak link.
   }
   \label{fig:sketch}
\end{figure}

If two full-shell nanowires with different mid radii $R_1$ and $R_2$ are brought together forming a S$_1$NS$_2$ JJ, see Fig. \ref{fig:sketch}, the jumps of fluxoid numbers $n_1$ and $n_2$ at each side occur at different values of the applied magnetic field, so that $n_1\neq n_2$ configurations can arise \footnote{Note that the winding number $n(\Phi)$ is only defined in the $S_1$ and $S_2$ regions (where it takes integer values $n_1$ and $n_2$, respectively), but not in the $N$ region, since the latter does not have a superconducting order parameter. Hence, in the split-shell geometry of Fig. \ref{fig:sketch} considered here, the geometric details of the $N$ region do not affect the $n_1, n_2$ values.}. We find that the superconducting critical current $I_c$ is finite when $n_1= n_2$, but it is blocked for magnetic field intervals for which $n_1\neq n_2$. 

The fluxoid valve effect can be understood using basic symmetry arguments. Far from the junction, the order parameter on side $\nu=1,2$ as a function of polar angle $\varphi$ is $\Delta_\nu(\varphi) = \Delta_\nu(0)e^{in_{\nu}\varphi}$, assuming cylindrical symmetry of the hybrid wire bulk. If $n_1\neq n_2$, a phase bias $\phi$ applied to the junction can be represented by changing this bulk boundary condition from $\Delta_\nu(\varphi)$ to $\Delta_\nu(\varphi+\delta\varphi)$, where $\delta\varphi$ is related to $\phi$ through $(n_1-n_2)\delta\varphi = \phi$. This transformation then relates the phase bias $\phi$ to a finite rotation $\delta \varphi$ of the boundary conditions around the nanowire axis. If the system as a whole (bulk nanowires and junction region) preserves cylindrical symmetry, the total free energy $F$ remains unchanged by this rotation, so that $\partial_\phi F(\phi) =0$, and hence the Josephson current $I(\phi) = (2e/\hbar)\partial_\phi F(\phi)$ and the critical current $I_c = \max_\phi |I(\phi)|$ will be zero. If $n_1=n_2$, a phase bias cannot be reduced to a finite rotation, and the Josephson current is finite in general.

We confirm these symmetry-based expectations via numerical simulations. We show that fluxoid valves with cylindrical symmetry allow perfect blocking of the supercurrent. 

If the cylindrical symmetry is broken, a finite supercurrent leakage may develop for $n_1\neq n_2$. We characterize the fidelity of the valve effect with a fluxoid-valve quality factor $Q^{FV}\in[0,1]$ that depends on the fluxoid numbers and the degree of symmetry breaking at the junction.

Due to the combination of Rashba spin-orbit coupling (SOC) in the semiconductor core and superconducting proximity effect from the shell, full-shell hybrid nanowires can host a topological superconducting phase \cite{Vaitiekenas:S20,*Vaitiekenas:ErrS25} for certain nanowire parameters if $B$ is such that an odd number of fluxoids thread the wire \footnote{Note that topological superconductivity can also be found in even number LP lobes if the cylindrical symmetry of the hybrid wire is sufficiently broken~\cite{Vaitiekenas:S20,*Vaitiekenas:ErrS25,Paya:PRB24}.}. In such a case, Majorana zero modes (MZMs) appear at the hybrid nanowire ends. In this work, we consider topological JJs with MZMs as well as junctions in the topologically-trivial regime (which we dub trivial regime in what follows). By analyzing both regimes, we find that the presence of MZMs in topological JJs may enhance substantially the fluxoid-valve quality factor.

Finally, it is worth noting that the Josephson current blocking for different fluxoids discussed here is related to a similar effect studied over the years for JJs in coaxially cylindrical and planar Corbino geometries \cite{Tilley:PL66,Sherrill:PRB79,Bhushan:PB81,Sherrill:PLA81,Burt:PLA81,Wang:JLTP91,Hadfield:PRB03,Clem:PRB10,Zhang:CPB22}. However, while prior works deal with radial transport, our valve effect concerns longitudinal transport \emph{along} the wire axis. This opens up new and useful possibilities, such as more flexible connectivity in nanowire circuits, or the possibility to tune the valve through a local gate at the junction (which can modify not only its transparency but also the valve symmetry). Moreover, the combination of hybrid superconductor-semiconductor nanostructures with SOC opens the possibility of a \emph{topological} fluxoid valve effect, a scenario that has not been explored before.

\section{Fluxoid valve in the topologically trivial regime}
\label{trivial}

We consider a short (normal length $L_N\rightarrow 0$) JJ like the one in Fig. \ref{fig:sketch}. $R_1$ and $R_2$ are different, but for simplicity we take the rest of the parameters the same, and S$_1$ and S$_2$ are semi-infinite. We further assume that the SOC in the core is negligible, $\alpha\rightarrow 0$, so that the superconducting sections are automatically in the trivial regime. Note that this assumption is not necessary. We obtain very similar results for $\alpha\neq 0$ as long as the wires remain in the trivial regime, i.e., for parameters in the trivial regions of the topological phase diagram. The reason is that the SOC has only a weak effect on the subgap states \cite{Paya:PRB24} and thus on the short-junction $I_c$ \cite{Paya:PRB25}, unless the system enters the topological regime and MZMs appear (as we will discuss in the next section).

To model the system, we use a microscopic approach based on Green's functions, which was recently described in Ref. \cite{Paya:PRB25}. There, the Josephson effect between equal left and right superconducting sections was analyzed in detail, finding a rich phenomenology for the critical current $I_c$ and the current-phase relation that depends on the semiconductor core model and system parameters. These parameters are geometrical (shell mid radius $R$ and thickness $d$), material (superconductor gap $|\Delta_0|$ and diffusive coherence length $\xi$ in the shell at $B=0$, core chemical potential $\mu$, SOC $\alpha$ and effective mass $m^*$) and structural (superconductor-semiconductor decay rate $\Gamma$ and junction normal transmission $T_{\rm N}$). 

\begin{figure}
   \centering
   \includegraphics[width=\columnwidth]{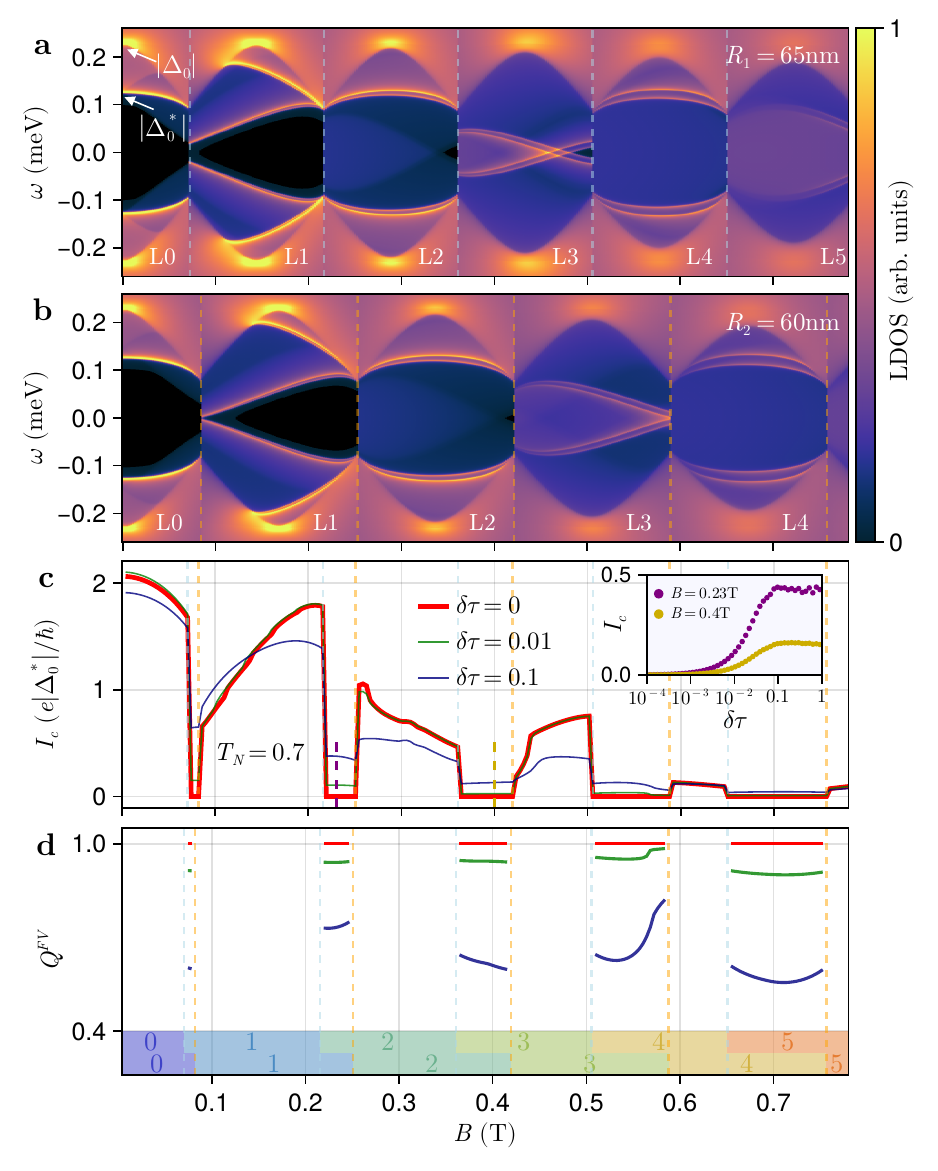}
   \caption{(a,b) Local density of states (LDOS) at the end of a semi-infinite full-shell hybrid nanowire with mid radius $R_1$ (a) and $R_2$ (b), as a function of energy $\omega$ and applied magnetic field $B$. Several Little-Parks (LP) lobes are displayed characterized by fluxoid numbers $n=1, 2, 3...$ (c) Critical current $I_c$ (normalized to the superconducting quantum unit~\cite{Beenakker:92} $e |\Delta_0^*|/\hbar$, where $|\Delta_0^*|$ is the induced gap at $B=0$) as a function of $B$ for a short junction with normal transmission $T_{\rm{N}}$. Red curve corresponds to a cylindrically symmetric junction. A perfect valve effect is achieved with $I_c=0$ in $B$ intervals $B^{n_1}_{n_2}$ where $n_1\neq n_2$. The valve effect is partially lifted (green and blue curves) for junctions with broken cylindrical symmetry, characterized by mode mixing parameter $\delta\tau$. Inset: $I_c$ vs  $\delta\tau$ for two different values of $B$. (d) Fluxoid-valve quality factor $Q^{\rm{FV}}$ versus $B^{n_1}_{n_2}$ for different values of $\delta\tau$. Parameters: $R_1=65$nm, $R_2=60$nm, $W = 20$nm, $d\rightarrow 0$, $|\Delta_0| = 0.23$meV, $\xi_1=\xi_2 = 70$nm, $m^*=0.023 m_e$, $\alpha\rightarrow 0$,  $\mu = 0.5$meV, $g = 1$, $\Gamma = |\Delta_0|$, $T_{\rm{N}}=0.7$ and $a_0 = 5$nm.}
   \label{fig:triv}
\end{figure}

The valve effect described here occurs independently of the core model, since it is a consequence of symmetry arguments. However, the quantitative values of $I_c$ for $n_1=n_2$ depend on the microscopic details of the core and the $N$ region. For the sake of simplicity we consider a short junction and a tubular model for the core, described in detail in Appendix A of Ref. \cite{Paya:PRB24a}, where the core charge is concentrated in a tube of thickness $W$; see Fig. \ref{fig:sketch}. This is a good model for realistic hybrid nanowires with an Ohmic interface between the shell and the core, where a charge accumulation layer typically develops in the core close to the superconductor-semiconductor interface. The diffusive superconductor shell is introduced as a self energy on the core surface, whose dependence with flux is computed self-consistently. We perform tight-binding numerical simulations of the proximitized core at zero temperature, with discretization lattice parameter $a_0$ in the $z$ direction. We specialize to Al/InAs parameters, although other III-V nanowires interfaced with $s$-wave superconductors should give similar results.

In the first two rows of Fig. \ref{fig:triv} we show the local density of states (LDOS) at the end of a single semi-infinite full-shell nanowire of radius $R_1$ (a) and $R_2$ (b) versus magnetic field $B$. Without loss of generality, we take $R_1>R_2$. The spectrum of these wires is characterized by the LP effect of the tubular shell, by which the shell gap is modulated with $B$ in a series of \emph{lobes} characterized by an integer number $n$ of fluxoids \cite{Vaitiekenas:PRB20,Sabonis:PRL20,Vekris:SR21}. Due to the different radii, five LP lobes are visible in (a) for the displayed magnetic field range, whereas only four in (b). Below the shell gap edge, there appear a number of Andreev states in the semiconductor core, called Caroli-de Gennes-Matricon (CdGM) analogs \cite{San-Jose:PRB23,Paya:PRB24}, which have been recently observed experimentally \cite{Deng:PRL25}. They are shell-induced Van Hove singularities in populated core subbands.

The critical current versus magnetic field is presented in Fig. \ref{fig:triv}(c). For a cylindrically symmetric S$_1$NS$_2$ junction (red curve), $I_c$ presents a perfect valve effect wherein the total Josephson current (and thus also $I_c$) is exactly zero for the $B$ intervals for which $n_1\neq n_2$, which we call $B^{n_1}_{n_2}$. In contrast, $I_c$ is finite (and proportional to $T_N$~\cite{Paya:PRB25}) for the $B$ intervals for which $n_1=n_2$, called $B^{n_1}_{n_1}$. The flux dependence of $I_c$ in these intervals is complicated and depends on the LP modulation of the shell gap, as well as the behavior of the CdGM analogs with flux \cite{Paya:PRB25} in the two S$_{1,2}$ sections. Note that $I_c$ does not only depend on the subgap spectrum, but also on the continuum above the gap that is included in our Green's function formulation. The latter contribution cannot be neglected, even in the short-junction limit considered here, since the Andreev approximation is not satisfied in nanowires close to depletion~\cite{Beenakker:PRL91,Beenakker:NaMS92, Setiawan:PRR22,Kruti:PRB24}. 

The evolution of $I_c$ with flux depends qualitatively on the ratio of the radii $R_1$ and $R_2$ at either side of the junction. In Fig. \ref{fig:triv}, we have chosen a case where $R_1<\sqrt{2}R_2$ (with $R_1>R_2)$. In this range, the valve closes within $B$ intervals given by $\frac{n_1 \Phi_0}{2\pi R_1^2} < B^{n_1}_{n_2} < \frac{n_1 \Phi_0}{2\pi R_2^2}$, with $n_{1,2}$ being contiguous $\mathbb{N}$ numbers. These intervals increase with $n_1$, see Fig. \ref{fig:triv}(c), and, eventually, the valve closes definitively for $B > \frac{\Phi_0}{2\pi \left(R_2^2- R_1^2\right)}$. In contrast, if $R_1 \geq \sqrt{2} R_2$, the valve remains closed already for $B > \frac{\Phi_0}{2 \pi R_1^2}$, and there are no $I_c$ revivals after $B^0_0$.

We now consider JJs with broken cylindrical symmetry. The asymmetry may come from a (sufficiently) non-circular cross section of the nanowires, or from a normal junction with broken symmetry due for example to the presence of a perpendicular electric field, like the one created by a plunger gate used to tune the junction transparency. We model this latter case in terms of a dimensionless parameter $\delta\tau\in[0,1]$ that quantifies the degree of junction asymmetry and that, in practice, introduces transverse mode mixing at the JJ; see details of the model in Appendix \ref{Ap:mm}. In this case, a small critical current develops for $n_1\neq n_2$; green and blue curves in Fig. \ref{fig:triv}(c) show results for two different $\delta\tau$ values. In the inset we show how $I_c$ reemerges from zero for a closed-valve $B$ (i.e., with $n_1\neq n_2$) as $\delta\tau$ is increased. Therefore, the fluxoid valve can be operated either by changing the magnetic field or by means of a local gate at the junction that gradually breaks the system's cylindrical symmetry.

We characterize the fidelity of the valve by defining a fluxoid-valve quality factor
\beq
Q^{\rm{FV}}=\frac{I_c^{(n_1=n_2)}-I_c^{(n_1\neq n_2)}}{I_c^{(n_1=n_2)}},
\label{QF}
\eeq
where $I_c^{(n_1=n_2)}$ and $I_c^{(n_1\neq n_2)}$ are critical currents within $B^{n_1}_{n_1}$ and $B^{n_1}_{n_2}$, respectively. We choose $B^{n_1}_{n_1}$ instead of $B^{n_2}_{n_2}$ as the reference, since $I_c$ tends to be larger in the former for $n_1<n_2$. Note that $Q^{\rm{FV}}\in [0,1]$. To represent $Q^{\rm{FV}}$ we fix $I_c^{(n_1=n_2)}$ to the rightmost edge of $B^{n_1}_{n_1}$ and compute $I_c^{(n_1\neq n_2)}$ as a function of flux across $B^{n_1}_{n_2}$. The results are shown in Fig. \ref{fig:triv}(d). $Q^{\rm{FV}}$ is perfect (i.e., 1) for a cylindrically symmetric junction, and decreases as $\delta\tau$ increases. 

In Fig. \ref{fig:triv} we have considered full-shell nanowires in the nondestructive LP regime, meaning that the shell gap never closes between lobes for both the S$_1$ and S$_2$ sections. For thin $d\rightarrow 0$ shells this happens for $R_\nu\gtrsim  0.6\xi_\nu$ ~\cite{San-Jose:PRB23} (a more general condition for $d\neq 0$ depends on the applied field). For narrower wires the superconducting sections may enter the destructive LP regime \cite{Schwiete:PRB10}, characterized by a full collapse of the superconducting gap within finite $B$ intervals between lobes. For $B$ values for which the gap in one (or both) of the hybrid wires collapses, $I_c$ vanishes trivially, simply because the shell becomes a normal metal and can no longer support a Josephson current. This kind of supercurrent suppression has nothing to do with the fluxoid valve effect (it also happens without a fluxoid mismatch), and it is discussed in Appendix \ref{Ap:destructive}.

\section{Fluxoid valve in the topological regime}
\label{topological}

\begin{figure}
   \centering
   \includegraphics[width=\columnwidth]{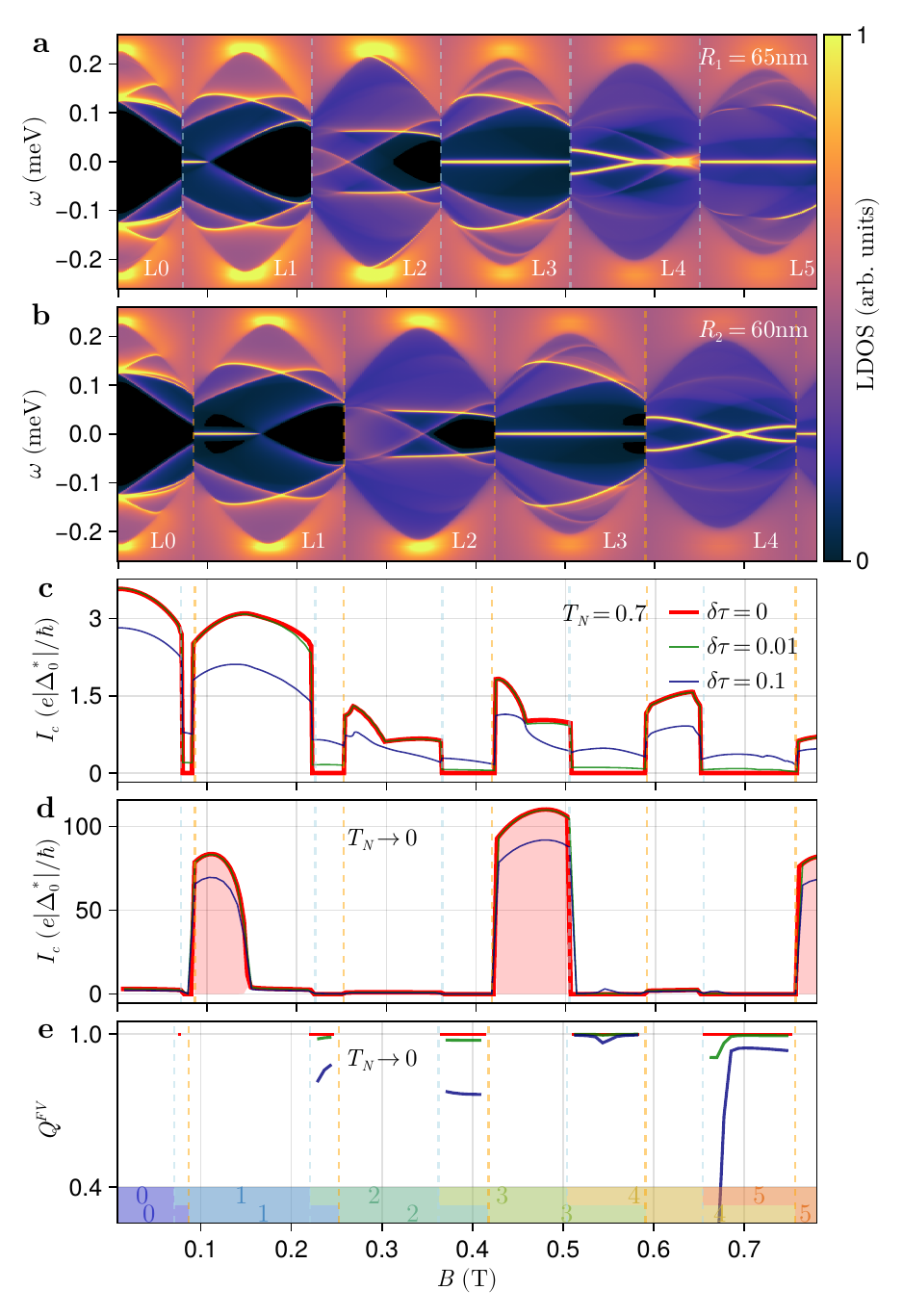}
   \caption{Same as Fig. \ref{fig:triv} but for a JJ with full-shell hybrid nanowires that can enter the topological superconducting regime. Majorana zero-energy peaks appear in the LDOS of (a,b) for certain flux intervals in odd LP lobes. For high junction transparency (c), $I_c$ is very similar to the trivial JJ case. However, for tunnel JJs (d), $I_c$ is affected by the presence of MZMs (whose contribution is highlighted with red shaded areas). Parameters as in Fig. \ref{fig:triv} except for $\alpha = 20$~meVnm and $g = 0$. 
   }
   \label{fig:topo}
\end{figure}

In Fig. \ref{fig:topo} we consider the fluxoid valve effect for a topological S$_1$NS$_2$ short JJ for semi-infinite superconducting sections. All system parameters are the same as in Fig. \ref{fig:triv} except for the values of SOC, chemical potential and $g$-factor, which are now selected so that one or both hybrid nanowires can enter the topological regime. This means that MZMs appear at the junction edges, which are visible as zero-energy peaks in LDOS for certain B intervals within odd LP lobes; see Figs. \ref{fig:topo}(a,b). \label{Ap:topo}

The presence of MZMs at the junction has a small impact on $I_c$ at high junction transparency; see  Fig. \ref{fig:topo}(c). The MZM contribution to $I_c$ is highlighted by a red shade. It is present only for B intervals for which there are Majorana zero-energy peaks in the LDOS of both superconducting sections S$_{1,2}$. In contrast, in the tunneling regime, the Majorana contribution is dominant; see Fig. \ref{fig:topo}(d). The reason is that $I_c$ scales as $\sqrt{T_N}$ for a topological JJ instead of linearly in $T_N$ like in the trivial regime. This behavior was discussed in Ref. \cite{Paya:PRB25}. Lastly, the fluxoid-valve quality factor $Q^{\rm{FV}}$ versus $B^{n_1}_{n_2}$ is shown in Fig. \ref{fig:topo}(e) for different values of the symmetry-breaking parameter $\delta\tau$ and for $T_N\rightarrow 0$. Its behavior is very similar to that of the trivial JJ, except for B intervals $B^{n_1}_{n_2}$ for which there are MZMs in both S$_{1,2}$ sections in $B^{n_1}_{n_1}$. In that case, $Q^{\rm{FV}}$ is almost perfect irrespective of $\delta\tau$; see the $B^4_3$ interval in Fig. \ref{fig:topo}(e).

In all our numerical simulations we have considered semi-infinite superconducting sections. We have checked, however, that finite-length hybrid nanowires give the same qualitative results for trivial junctions. We obtain only small quantitative differences for $I_c$ in the magnetic field intervals where $I_c$ is finite \cite{Paya:PRB25} (not shown). Finite wire lengths only have a relevant impact on $I_c$ for low-transparency topological junctions. In finite-length topological nanowires, MZMs appear also at the far nanowire ends. These outer MZMs can hybridize with the MZMs at the junction. When the intra-nanowire MZM splitting due to their finite overlap exceeds the MZM splitting across the junction, the low-transparency enhancement of $Q^{\rm{FV}}$ mentioned above is suppressed.

\section{Conclusions}
\label{conclusions}

We have introduced a supercurrent valve effect that is possible in S$_1$NS$_2$ JJs along the axial direction with doubly connected superconducting sections of different radii. This type of system has already been fabricated by growing shells around semiconductor nanowires with a spatially varying radius \cite{Nygard:}. 

Following the analogy with a spin valve in the introduction, in this case we have a non-superconducting material (the semiconductor junction) between two (hybrid nanowire) superconducting domains with different fluxoid ``pseudospin'' numbers $n_1$ and $n_2$. The transmission of a supercurrent across the junction depends on the relative fluxoid ``alignment'' (parallel or antiparallel) of the two superconductors, which can be controlled by the applied magnetic field. For ``parallel polarization", $n_1=n_2$, the supercurrent flows unhindered, while for an ``antiparallel polarization", $n_1\neq n_2$, the supercurrent is suppressed, even reaching zero if the wavefunctions of scattering modes remain cylindrically symmetric throughout the device.

There are, however, several important differences between a spin and a fluxoid valve. While the magnetic moment is a vector that can point in any direction spanned by the easy axes (one or several) of the device, the fluxoid number is always a scalar. As a consequence, thermal fluctuations do not negatively impact the quality factor of a fluxoid valve. This is unlike spin valves, where magnetic moment fluctuations can degrade the valve performance, even between half-metals \cite{Everschor-Sitte:JoAP14}.
Moreover, the fluxoid valve has many quantized ``pseudospin" states $n_{1,2}\in\mathbb{Z}$, not just a non-quantized magnetic moment $M$ and $-M$ along a given axis. Therefore, we can define different fluxoid valve quality factors
$Q^{\rm{FV}}$ for different $B$ intervals $B^{n_1}_{n_2}$ with contiguous integers $n_{1,2}$. Also, there is no fluxoid analog of minority spins in spin valves, which also degrade magnetoresistance.

Due to the combination of SOC in the semiconductor core and superconducting proximity effect from the shell, full-shell hybrid nanowires can host a topological superconducting phase for certain nanowire parameters and $B$ intervals. In such a case, MZMs appear at the junction edges. We have also analyzed the fluxoid valve effect for topological superconducting sections. We have found that for high transparency JJs, the presence of Majoranas does not significantly affect the critical current. However, for tunnel junctions, MZMs can significantly enhance the on-off valve ratio.  

All the numerical codes used in this paper were based on the Quantica.jl package \cite{San-Jose:25a}. The specific code to build the nanowire Hamiltonian and to perform and plot the calculations is available at Refs. \cite{Paya:25a} and \cite{Paya:25h}, respectively. Visualizations were made with the Makie.jl package \cite{Danisch:JOSS21}. 

\acknowledgments
We thank Jesper Nygård for stimulating discussions. This research was supported by Grants MDM-2014-0377, FPU20/01871, PID2021-122769NB-I00, PID2021-125343NB-I00, TED2021-130292B-C41, PRE2022-101362, PID2023-150224NB-I00 and CEX2024-001445-S, funded by MICIU/AEI/10.13039/501100011033, ``ERDF A way of making Europe'' and ``ESF+''. Part of this research project was made possible through the access granted by the Galician Supercomputing Center (CESGA) to its supercomputing infrastructure. 

\appendix

\section{Mode mixing model}
\label{Ap:mm}

The two full-shell nanowires at either side of the JJ are modeled with a Bogoliubov-de Gennes (BdG) Hamiltonian, $H_{\rm BdG}$ \cite{De-Gennes:18,Tinkham:96}. The possible forms of this model and their microscopic justification were extensively discussed in Refs. \cite{San-Jose:PRB23,Paya:PRB24, Paya:PRB24a}. For the purpose of this work, the essential property of $H_{\rm BdG}$ is its cylindrical symmetry. This has generic implications about the quantum numbers of its eigenstates. \label{Ap:mm}

The operator $J_z = -i\partial_\varphi + \frac{1}{2}\sigma_z + \frac{1}{2}n \tau_z$ is a generalized angular momentum that satisfies $\left[J_z, H_{\rm BdG}\right] = 0$, even in the presence of a finite fluxoid \cite{Vaitiekenas:S20, *Vaitiekenas:ErrS25}. Here $\sigma_z$ ($\tau_z$) is the third Pauli matrix in spin (electron-hole) space and $n$ is the fluxoid threading the wire. Consequently, the eigenstates of $H_{\rm BdG}$ can be labeled by the $J_z$'s quantum number $m_J$,
\begin{equation}
    \Psi_{m_J}(\vec{r}) \propto e^{i\left(m_J - \frac{1}{2}\sigma_z -\frac{1}{2}n\tau_z\right)\varphi} \Psi_{m_J}(r, z),
\end{equation}
where $\Psi_{m_J}$ is a Nambu spinor in the basis $\Psi = \left(\psi_\uparrow, \psi_\downarrow, \psi_\downarrow^\dagger, -\psi_\uparrow^\dagger\right)^{\rm T}$. Since $n \in \mathbb{Z}$, $m_J$ is constrained to
\begin{equation}
    m_J \in \left\{\begin{matrix}
        \mathbb{Z} + \frac{1}{2} & n\text{ even} \\
        \mathbb{Z} & n  \text{ odd}
    \end{matrix}\right. .
\end{equation}

The above applies to each of the two decoupled full-shell hybrid nanowires S$_1$ and S$_2$. The JJ itself is modeled by a hopping $V(r, \varphi) = V(r) V(\varphi)$ between sites on either side of the junction at radius $r$ and polar angle $\varphi$. The matrix elements of the hopping across the junction read
\begin{eqnarray}
    &&\mel{\Psi_{m_J^{(1)}}}{V}{\Psi_{m_J^{(2)}}} \propto \nonumber\\
    &&\int_0^{2\pi} e^{i\left[\left(m_J^{(2)}-m_J^{(1)}\right) -\frac{1}{2}\left(n_2-n_1\right) \tau_z\right]\varphi}V(\varphi)d\varphi.
    \label{eq:V12}
\end{eqnarray}
Expanding $V(\varphi)$ in a Fourier series,
\begin{equation}
    V(\varphi) = \sum_\ell V_\ell e^{i \ell \varphi},
\end{equation}
with $\ell$ the (integer) harmonic index and $V_\ell$ its amplitude, leads to a selection rule for the coupling between generalized angular momentum modes of the two superconducting sections. If $n_1 = n_2$, $V_\ell$ couples only $m_J$'s that obey the selection rule
\begin{equation}
    m_J^{(2)} - m_J^{(1)} = - \ell.
\end{equation}
However, if $n_1 \neq n_2$, the matrix structure of the coupling in Eq. \eqref{eq:V12} is more complex and depends on the electron-hole character,
\begin{equation}
    m_J^{(2)} - m_J^{(1)} = \frac{1}{2}\left(n_2 - n_1\right)\tau_z - \ell.
\end{equation}

In the main text, we considered both a cylindrically symmetric ($\ell = 0$) junction, $V(\varphi) = V$, and a junction that includes also a dipolar ($\ell = \pm 1$) perturbation,
\begin{equation}
    V(\varphi) = V\left[1 + \delta\tau \cos(\varphi) \right].
\end{equation}
This perturbation (controlled by the dimensionless parameter $\delta \tau \in [0, 1]$) models the electric potential from a gate electrode below the junction that pushes the electron flow in the junction towards angles around $\varphi = 0$. 

\section{Destructive LP regime}
\label{Ap:destructive}

\begin{figure}
   \centering
   \includegraphics[width=\columnwidth]{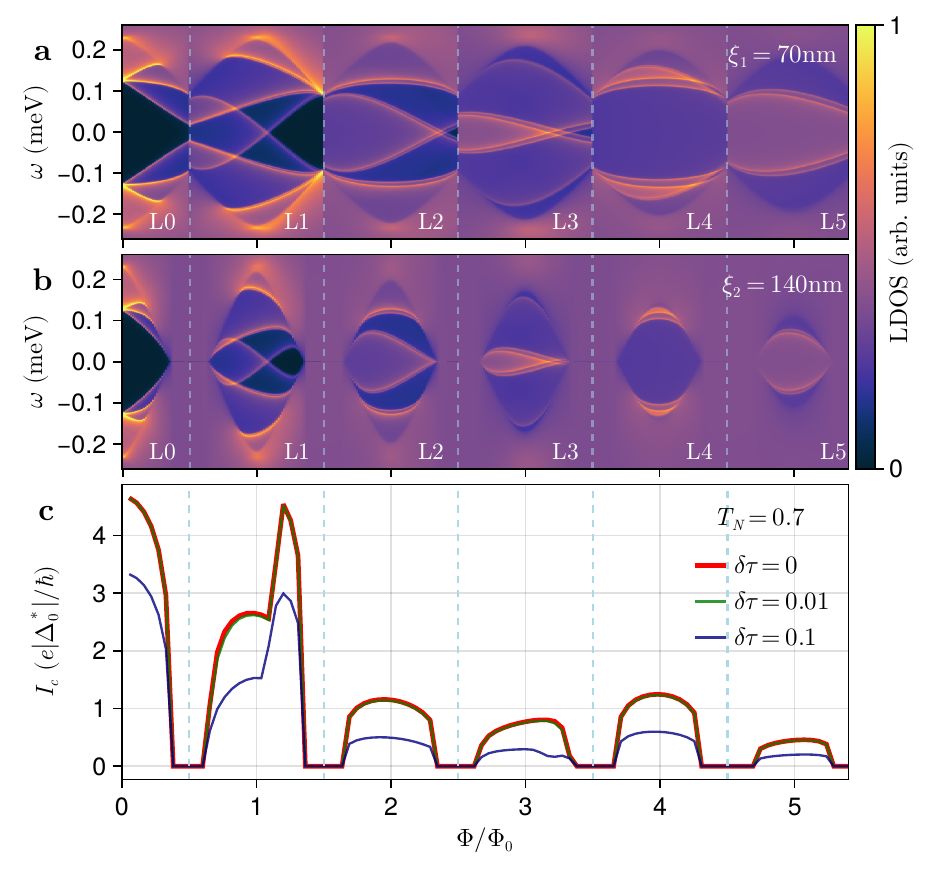}
   \caption{Same as Fig. \ref{fig:triv}(a-c) but for superconducting sections with the same radius and with different diffusive coherence lengths $\xi_1\neq\xi_2$. S$_1$ is in the nondestructive LP regime, whereas S$_2$ is in the destructive one. Parameters as in Fig. \ref{fig:triv} except for $R_1=R_2=65$nm, $\xi_1=70$nm, $\xi_2=140$nm, and $\mu = 2$~meV.}
   \label{fig:xi}
\end{figure}

In the main text we have studied full-shell nanowires for which both the S$_1$ and S$_2$ sections are in the nondestructive LP regime (corresponding to $R_\nu\gtrsim 0.6\xi_\nu$ for $d\rightarrow 0$). If we decrease any of the hybrid nanowire mid radii while keeping the diffusive coherence length unchanged, that superconducting section will eventually enter the destructive LP regime \cite{Schwiete:PRB10}. Then, its superconducting gap will vanish within finite $B$ intervals between lobes, so that $I_c$ also vanishes. Such $I_c$ suppression has a trivial origin, since in those situations the JJ has simply become normal on one (or both) sides and cannot support a Josephson current. This phenomenon is unrelated to the fluxoid valve effect and actually occurs also for equal-radius superconducting sections with $n_1=n_2$.

We illustrate this scenario in Fig. \ref{fig:xi} for a JJ in the trivial regime. To avoid confounding the fluxoid-valve and the destructive-LP-regime supercurrent blockings, we fix $R_1=R_2$ (no valve effect), but we allow $\xi_1\neq\xi_2$ for generality, so that the LP effect in $S_1$ is nondestructive, but in $S_2$ it is destructive. The LDOS of the two semi-infinite superconducting sections are shown in Figs. \ref{fig:xi}(a,b), respectively, and $I_c$ versus $B$ in Fig. \ref{fig:xi}(c). Note that $I_c=0$ in the $B$ intervals for which the gap in section S$_2$ collapses, regardless of the cylindrical symmetry breaking.

In a general situation where one or the two superconducting sections are in the destructive LP regime and also $R_1\neq R_2$, there will be $B$ intervals for which the system displays a valve effect, coexisting with other $B$ intervals for which $I_c=0$ simply because there is no JJ anymore. One can envision several ways to distinguish between these two scenarios experimentally. One possibility is to change the junction gate potential at a fixed magnetic field and check whether the critical current deviates from strictly zero as one breaks the junction cylindrical symmetry (note that only the fluxoid valve effect is lifted by symmetry breaking). Another possibility is to attach external tunnel probes to the two superconducting shells to detect whether both display a finite gap for that $B$ field. 

\bibliography{Josephson}

\begin{thebibliography}{73}%
\makeatletter
\providecommand \@ifxundefined [1]{%
 \@ifx{#1\undefined}
}%
\providecommand \@ifnum [1]{%
 \ifnum #1\expandafter \@firstoftwo
 \else \expandafter \@secondoftwo
 \fi
}%
\providecommand \@ifx [1]{%
 \ifx #1\expandafter \@firstoftwo
 \else \expandafter \@secondoftwo
 \fi
}%
\providecommand \natexlab [1]{#1}%
\providecommand \enquote  [1]{``#1''}%
\providecommand \bibnamefont  [1]{#1}%
\providecommand \bibfnamefont [1]{#1}%
\providecommand \citenamefont [1]{#1}%
\providecommand \href@noop [0]{\@secondoftwo}%
\providecommand \href [0]{\begingroup \@sanitize@url \@href}%
\providecommand \@href[1]{\@@startlink{#1}\@@href}%
\providecommand \@@href[1]{\endgroup#1\@@endlink}%
\providecommand \@sanitize@url [0]{\catcode `\\12\catcode `\$12\catcode `\&12\catcode `\#12\catcode `\^12\catcode `\_12\catcode `\%12\relax}%
\providecommand \@@startlink[1]{}%
\providecommand \@@endlink[0]{}%
\providecommand \url  [0]{\begingroup\@sanitize@url \@url }%
\providecommand \@url [1]{\endgroup\@href {#1}{\urlprefix }}%
\providecommand \urlprefix  [0]{URL }%
\providecommand \Eprint [0]{\href }%
\providecommand \doibase [0]{https://doi.org/}%
\providecommand \selectlanguage [0]{\@gobble}%
\providecommand \bibinfo  [0]{\@secondoftwo}%
\providecommand \bibfield  [0]{\@secondoftwo}%
\providecommand \translation [1]{[#1]}%
\providecommand \BibitemOpen [0]{}%
\providecommand \bibitemStop [0]{}%
\providecommand \bibitemNoStop [0]{.\EOS\space}%
\providecommand \EOS [0]{\spacefactor3000\relax}%
\providecommand \BibitemShut  [1]{\csname bibitem#1\endcsname}%
\let\auto@bib@innerbib\@empty
\bibitem [{\citenamefont {Haselman}\ and\ \citenamefont {Hauck}(2010)}]{Haselman:PI10}%
  \BibitemOpen
  \bibfield  {author} {\bibinfo {author} {\bibfnamefont {M.}~\bibnamefont {Haselman}}\ and\ \bibinfo {author} {\bibfnamefont {S.}~\bibnamefont {Hauck}},\ }\bibfield  {title} {\bibinfo {title} {The {{Future}} of {{Integrated Circuits}}: {{A Survey}} of {{Nanoelectronics}}},\ }\href {https://doi.org/10.1109/JPROC.2009.2032356} {\bibfield  {journal} {\bibinfo  {journal} {Proceedings of the IEEE}\ }\textbf {\bibinfo {volume} {98}},\ \bibinfo {pages} {11} (\bibinfo {year} {2010})}\BibitemShut {NoStop}%
\bibitem [{\citenamefont {Parkin}\ \emph {et~al.}(2003)\citenamefont {Parkin}, \citenamefont {Jiang}, \citenamefont {Kaiser}, \citenamefont {Panchula}, \citenamefont {Roche},\ and\ \citenamefont {Samant}}]{Parkin:PI03}%
  \BibitemOpen
  \bibfield  {author} {\bibinfo {author} {\bibfnamefont {S.}~\bibnamefont {Parkin}}, \bibinfo {author} {\bibfnamefont {X.}~\bibnamefont {Jiang}}, \bibinfo {author} {\bibfnamefont {C.}~\bibnamefont {Kaiser}}, \bibinfo {author} {\bibfnamefont {A.}~\bibnamefont {Panchula}}, \bibinfo {author} {\bibfnamefont {K.}~\bibnamefont {Roche}},\ and\ \bibinfo {author} {\bibfnamefont {M.}~\bibnamefont {Samant}},\ }\bibfield  {title} {\bibinfo {title} {Magnetically engineered spintronic sensors and memory},\ }\href {https://doi.org/10.1109/JPROC.2003.811807} {\bibfield  {journal} {\bibinfo  {journal} {Proceedings of the IEEE}\ }\textbf {\bibinfo {volume} {91}},\ \bibinfo {pages} {661} (\bibinfo {year} {2003})}\BibitemShut {NoStop}%
\bibitem [{\citenamefont {{\v Z}uti{\'c}}\ \emph {et~al.}(2004)\citenamefont {{\v Z}uti{\'c}}, \citenamefont {Fabian},\ and\ \citenamefont {Das~Sarma}}]{Zutic:RMP04}%
  \BibitemOpen
  \bibfield  {author} {\bibinfo {author} {\bibfnamefont {I.}~\bibnamefont {{\v Z}uti{\'c}}}, \bibinfo {author} {\bibfnamefont {J.}~\bibnamefont {Fabian}},\ and\ \bibinfo {author} {\bibfnamefont {S.}~\bibnamefont {Das~Sarma}},\ }\bibfield  {title} {\bibinfo {title} {Spintronics: {{Fundamentals}} and applications},\ }\href {https://doi.org/10.1103/RevModPhys.76.323} {\bibfield  {journal} {\bibinfo  {journal} {Rev. Mod. Phys.}\ }\textbf {\bibinfo {volume} {76}},\ \bibinfo {pages} {323} (\bibinfo {year} {2004})}\BibitemShut {NoStop}%
\bibitem [{\citenamefont {Niedzielski}\ \emph {et~al.}(2018)\citenamefont {Niedzielski}, \citenamefont {Bertus}, \citenamefont {Glick}, \citenamefont {Loloee}, \citenamefont {Pratt},\ and\ \citenamefont {Birge}}]{Niedzielski:PRB18}%
  \BibitemOpen
  \bibfield  {author} {\bibinfo {author} {\bibfnamefont {B.~M.}\ \bibnamefont {Niedzielski}}, \bibinfo {author} {\bibfnamefont {T.~J.}\ \bibnamefont {Bertus}}, \bibinfo {author} {\bibfnamefont {J.~A.}\ \bibnamefont {Glick}}, \bibinfo {author} {\bibfnamefont {R.}~\bibnamefont {Loloee}}, \bibinfo {author} {\bibfnamefont {W.~P.}\ \bibnamefont {Pratt}},\ and\ \bibinfo {author} {\bibfnamefont {N.~O.}\ \bibnamefont {Birge}},\ }\bibfield  {title} {\bibinfo {title} {Spin-valve {{Josephson}} junctions for cryogenic memory},\ }\href {https://doi.org/10.1103/PhysRevB.97.024517} {\bibfield  {journal} {\bibinfo  {journal} {Phys. Rev. B}\ }\textbf {\bibinfo {volume} {97}},\ \bibinfo {pages} {024517} (\bibinfo {year} {2018})}\BibitemShut {NoStop}%
\bibitem [{\citenamefont {{Stoddart-Stones}}\ \emph {et~al.}(2022)\citenamefont {{Stoddart-Stones}}, \citenamefont {Montiel}, \citenamefont {Blamire},\ and\ \citenamefont {Robinson}}]{Stoddart-Stones:CP22}%
  \BibitemOpen
  \bibfield  {author} {\bibinfo {author} {\bibfnamefont {B.}~\bibnamefont {{Stoddart-Stones}}}, \bibinfo {author} {\bibfnamefont {X.}~\bibnamefont {Montiel}}, \bibinfo {author} {\bibfnamefont {M.~G.}\ \bibnamefont {Blamire}},\ and\ \bibinfo {author} {\bibfnamefont {J.~W.~A.}\ \bibnamefont {Robinson}},\ }\bibfield  {title} {\bibinfo {title} {Competition between the superconducting spin-valve effect and quasiparticle spin-decay in superconducting spin-valves},\ }\href {https://doi.org/10.1038/s42005-022-01003-0} {\bibfield  {journal} {\bibinfo  {journal} {Commun Phys}\ }\textbf {\bibinfo {volume} {5}},\ \bibinfo {pages} {1} (\bibinfo {year} {2022})}\BibitemShut {NoStop}%
\bibitem [{\citenamefont {Bobkov}\ \emph {et~al.}(2024)\citenamefont {Bobkov}, \citenamefont {Gordeeva}, \citenamefont {Johnsen~Kamra}, \citenamefont {Chourasia}, \citenamefont {Bobkov}, \citenamefont {Kamra},\ and\ \citenamefont {Bobkova}}]{Bobkov:PRB24}%
  \BibitemOpen
  \bibfield  {author} {\bibinfo {author} {\bibfnamefont {G.~A.}\ \bibnamefont {Bobkov}}, \bibinfo {author} {\bibfnamefont {V.~M.}\ \bibnamefont {Gordeeva}}, \bibinfo {author} {\bibfnamefont {L.}~\bibnamefont {Johnsen~Kamra}}, \bibinfo {author} {\bibfnamefont {S.}~\bibnamefont {Chourasia}}, \bibinfo {author} {\bibfnamefont {A.~M.}\ \bibnamefont {Bobkov}}, \bibinfo {author} {\bibfnamefont {A.}~\bibnamefont {Kamra}},\ and\ \bibinfo {author} {\bibfnamefont {I.~V.}\ \bibnamefont {Bobkova}},\ }\bibfield  {title} {\bibinfo {title} {Superconducting spin valves based on antiferromagnet/superconductor/antiferromagnet heterostructures},\ }\href {https://doi.org/10.1103/PhysRevB.109.184504} {\bibfield  {journal} {\bibinfo  {journal} {Phys. Rev. B}\ }\textbf {\bibinfo {volume} {109}},\ \bibinfo {pages} {184504} (\bibinfo {year} {2024})}\BibitemShut {NoStop}%
\bibitem [{\citenamefont {Wu}\ \emph {et~al.}(2018)\citenamefont {Wu}, \citenamefont {Huang}, \citenamefont {Fang}, \citenamefont {Yang}, \citenamefont {Wan}, \citenamefont {Yu}, \citenamefont {Feng}, \citenamefont {Wei},\ and\ \citenamefont {Han}}]{Wu:PRL18}%
  \BibitemOpen
  \bibfield  {author} {\bibinfo {author} {\bibfnamefont {H.}~\bibnamefont {Wu}}, \bibinfo {author} {\bibfnamefont {L.}~\bibnamefont {Huang}}, \bibinfo {author} {\bibfnamefont {C.}~\bibnamefont {Fang}}, \bibinfo {author} {\bibfnamefont {B.~S.}\ \bibnamefont {Yang}}, \bibinfo {author} {\bibfnamefont {C.~H.}\ \bibnamefont {Wan}}, \bibinfo {author} {\bibfnamefont {G.~Q.}\ \bibnamefont {Yu}}, \bibinfo {author} {\bibfnamefont {J.~F.}\ \bibnamefont {Feng}}, \bibinfo {author} {\bibfnamefont {H.~X.}\ \bibnamefont {Wei}},\ and\ \bibinfo {author} {\bibfnamefont {X.~F.}\ \bibnamefont {Han}},\ }\bibfield  {title} {\bibinfo {title} {Magnon {{Valve Effect}} between {{Two Magnetic Insulators}}},\ }\href {https://doi.org/10.1103/PhysRevLett.120.097205} {\bibfield  {journal} {\bibinfo  {journal} {Phys. Rev. Lett.}\ }\textbf {\bibinfo {volume} {120}},\ \bibinfo {pages} {097205} (\bibinfo {year} {2018})}\BibitemShut {NoStop}%
\bibitem [{\citenamefont {{San-Jose}}\ \emph {et~al.}(2009)\citenamefont {{San-Jose}}, \citenamefont {Prada}, \citenamefont {McCann},\ and\ \citenamefont {Schomerus}}]{San-Jose:PRL09}%
  \BibitemOpen
  \bibfield  {author} {\bibinfo {author} {\bibfnamefont {P.}~\bibnamefont {{San-Jose}}}, \bibinfo {author} {\bibfnamefont {E.}~\bibnamefont {Prada}}, \bibinfo {author} {\bibfnamefont {E.}~\bibnamefont {McCann}},\ and\ \bibinfo {author} {\bibfnamefont {H.}~\bibnamefont {Schomerus}},\ }\bibfield  {title} {\bibinfo {title} {Pseudospin {{Valve}} in {{Bilayer Graphene}}: {{Towards Graphene-Based Pseudospintronics}}},\ }\href {https://doi.org/10.1103/PhysRevLett.102.247204} {\bibfield  {journal} {\bibinfo  {journal} {Phys. Rev. Lett.}\ }\textbf {\bibinfo {volume} {102}},\ \bibinfo {pages} {247204} (\bibinfo {year} {2009})}\BibitemShut {NoStop}%
\bibitem [{\citenamefont {Jung}\ \emph {et~al.}(2020)\citenamefont {Jung}, \citenamefont {Ryu}, \citenamefont {Shin}, \citenamefont {Sohn}, \citenamefont {Huh}, \citenamefont {Koch}, \citenamefont {Jozwiak}, \citenamefont {Rotenberg}, \citenamefont {Bostwick},\ and\ \citenamefont {Kim}}]{Jung:NM20}%
  \BibitemOpen
  \bibfield  {author} {\bibinfo {author} {\bibfnamefont {S.~W.}\ \bibnamefont {Jung}}, \bibinfo {author} {\bibfnamefont {S.~H.}\ \bibnamefont {Ryu}}, \bibinfo {author} {\bibfnamefont {W.~J.}\ \bibnamefont {Shin}}, \bibinfo {author} {\bibfnamefont {Y.}~\bibnamefont {Sohn}}, \bibinfo {author} {\bibfnamefont {M.}~\bibnamefont {Huh}}, \bibinfo {author} {\bibfnamefont {R.~J.}\ \bibnamefont {Koch}}, \bibinfo {author} {\bibfnamefont {C.}~\bibnamefont {Jozwiak}}, \bibinfo {author} {\bibfnamefont {E.}~\bibnamefont {Rotenberg}}, \bibinfo {author} {\bibfnamefont {A.}~\bibnamefont {Bostwick}},\ and\ \bibinfo {author} {\bibfnamefont {K.~S.}\ \bibnamefont {Kim}},\ }\bibfield  {title} {\bibinfo {title} {Black phosphorus as a bipolar pseudospin semiconductor},\ }\href {https://doi.org/10.1038/s41563-019-0590-2} {\bibfield  {journal} {\bibinfo  {journal} {Nat. Mater.}\ }\textbf {\bibinfo {volume} {19}},\ \bibinfo {pages} {277} (\bibinfo {year} {2020})}\BibitemShut {NoStop}%
\bibitem [{\citenamefont {Li}\ \emph {et~al.}(2018)\citenamefont {Li}, \citenamefont {Zhang}, \citenamefont {Yin}, \citenamefont {Zhang}, \citenamefont {Watanabe}, \citenamefont {Taniguchi}, \citenamefont {Liu},\ and\ \citenamefont {Zhu}}]{Li:S18}%
  \BibitemOpen
  \bibfield  {author} {\bibinfo {author} {\bibfnamefont {J.}~\bibnamefont {Li}}, \bibinfo {author} {\bibfnamefont {R.-X.}\ \bibnamefont {Zhang}}, \bibinfo {author} {\bibfnamefont {Z.}~\bibnamefont {Yin}}, \bibinfo {author} {\bibfnamefont {J.}~\bibnamefont {Zhang}}, \bibinfo {author} {\bibfnamefont {K.}~\bibnamefont {Watanabe}}, \bibinfo {author} {\bibfnamefont {T.}~\bibnamefont {Taniguchi}}, \bibinfo {author} {\bibfnamefont {C.}~\bibnamefont {Liu}},\ and\ \bibinfo {author} {\bibfnamefont {J.}~\bibnamefont {Zhu}},\ }\bibfield  {title} {\bibinfo {title} {A valley valve and electron beam splitter},\ }\href {https://doi.org/10.1126/science.aao5989} {\bibfield  {journal} {\bibinfo  {journal} {Science}\ }\textbf {\bibinfo {volume} {362}},\ \bibinfo {pages} {1149} (\bibinfo {year} {2018})}\BibitemShut {NoStop}%
\bibitem [{\citenamefont {Pesin}\ and\ \citenamefont {MacDonald}(2012)}]{Pesin:NM12}%
  \BibitemOpen
  \bibfield  {author} {\bibinfo {author} {\bibfnamefont {D.}~\bibnamefont {Pesin}}\ and\ \bibinfo {author} {\bibfnamefont {A.~H.}\ \bibnamefont {MacDonald}},\ }\bibfield  {title} {\bibinfo {title} {Spintronics and pseudospintronics in graphene and topological insulators},\ }\href {https://doi.org/10.1038/nmat3305} {\bibfield  {journal} {\bibinfo  {journal} {Nature Mater}\ }\textbf {\bibinfo {volume} {11}},\ \bibinfo {pages} {409} (\bibinfo {year} {2012})}\BibitemShut {NoStop}%
\bibitem [{\citenamefont {Bao}\ and\ \citenamefont {Zhou}(2020)}]{Bao:NM20}%
  \BibitemOpen
  \bibfield  {author} {\bibinfo {author} {\bibfnamefont {C.}~\bibnamefont {Bao}}\ and\ \bibinfo {author} {\bibfnamefont {S.}~\bibnamefont {Zhou}},\ }\bibfield  {title} {\bibinfo {title} {Black phosphorous for pseudospintronics},\ }\href {https://doi.org/10.1038/s41563-020-0614-y} {\bibfield  {journal} {\bibinfo  {journal} {Nat. Mater.}\ }\textbf {\bibinfo {volume} {19}},\ \bibinfo {pages} {263} (\bibinfo {year} {2020})}\BibitemShut {NoStop}%
\bibitem [{\citenamefont {Schaibley}\ \emph {et~al.}(2016)\citenamefont {Schaibley}, \citenamefont {Yu}, \citenamefont {Clark}, \citenamefont {Rivera}, \citenamefont {Ross}, \citenamefont {Seyler}, \citenamefont {Yao},\ and\ \citenamefont {Xu}}]{Schaibley:NRM16}%
  \BibitemOpen
  \bibfield  {author} {\bibinfo {author} {\bibfnamefont {J.~R.}\ \bibnamefont {Schaibley}}, \bibinfo {author} {\bibfnamefont {H.}~\bibnamefont {Yu}}, \bibinfo {author} {\bibfnamefont {G.}~\bibnamefont {Clark}}, \bibinfo {author} {\bibfnamefont {P.}~\bibnamefont {Rivera}}, \bibinfo {author} {\bibfnamefont {J.~S.}\ \bibnamefont {Ross}}, \bibinfo {author} {\bibfnamefont {K.~L.}\ \bibnamefont {Seyler}}, \bibinfo {author} {\bibfnamefont {W.}~\bibnamefont {Yao}},\ and\ \bibinfo {author} {\bibfnamefont {X.}~\bibnamefont {Xu}},\ }\bibfield  {title} {\bibinfo {title} {Valleytronics in {{2D}} materials},\ }\href {https://doi.org/10.1038/natrevmats.2016.55} {\bibfield  {journal} {\bibinfo  {journal} {Nat Rev Mater}\ }\textbf {\bibinfo {volume} {1}},\ \bibinfo {pages} {1} (\bibinfo {year} {2016})}\BibitemShut {NoStop}%
\bibitem [{\citenamefont {Izhak}\ \emph {et~al.}(2024)\citenamefont {Izhak}, \citenamefont {Karnieli}, \citenamefont {Yesharim}, \citenamefont {Tsesses},\ and\ \citenamefont {Arie}}]{Izhak:OLO24}%
  \BibitemOpen
  \bibfield  {author} {\bibinfo {author} {\bibfnamefont {S.}~\bibnamefont {Izhak}}, \bibinfo {author} {\bibfnamefont {A.}~\bibnamefont {Karnieli}}, \bibinfo {author} {\bibfnamefont {O.}~\bibnamefont {Yesharim}}, \bibinfo {author} {\bibfnamefont {S.}~\bibnamefont {Tsesses}},\ and\ \bibinfo {author} {\bibfnamefont {A.}~\bibnamefont {Arie}},\ }\bibfield  {title} {\bibinfo {title} {All-optical spin valve effect in nonlinear optics},\ }\href {https://doi.org/10.1364/OL.517794} {\bibfield  {journal} {\bibinfo  {journal} {Opt. Lett., OL}\ }\textbf {\bibinfo {volume} {49}},\ \bibinfo {pages} {1025} (\bibinfo {year} {2024})}\BibitemShut {NoStop}%
\bibitem [{\citenamefont {Goffman}\ \emph {et~al.}(2017)\citenamefont {Goffman}, \citenamefont {Urbina}, \citenamefont {Pothier}, \citenamefont {Nyg{\aa}rd}, \citenamefont {Marcus},\ and\ \citenamefont {Krogstrup}}]{Goffman:NJP17}%
  \BibitemOpen
  \bibfield  {author} {\bibinfo {author} {\bibfnamefont {M.~F.}\ \bibnamefont {Goffman}}, \bibinfo {author} {\bibfnamefont {C.}~\bibnamefont {Urbina}}, \bibinfo {author} {\bibfnamefont {H.}~\bibnamefont {Pothier}}, \bibinfo {author} {\bibfnamefont {J.}~\bibnamefont {Nyg{\aa}rd}}, \bibinfo {author} {\bibfnamefont {C.~M.}\ \bibnamefont {Marcus}},\ and\ \bibinfo {author} {\bibfnamefont {P.}~\bibnamefont {Krogstrup}},\ }\bibfield  {title} {\bibinfo {title} {Conduction channels of an {{InAs-Al}} nanowire {{Josephson}} weak link},\ }\href {https://doi.org/10.1088/1367-2630/aa7641} {\bibfield  {journal} {\bibinfo  {journal} {New J. Phys.}\ }\textbf {\bibinfo {volume} {19}},\ \bibinfo {pages} {092002} (\bibinfo {year} {2017})}\BibitemShut {NoStop}%
\bibitem [{\citenamefont {Tosi}\ \emph {et~al.}(2019)\citenamefont {Tosi}, \citenamefont {Metzger}, \citenamefont {Goffman}, \citenamefont {Urbina}, \citenamefont {Pothier}, \citenamefont {Park}, \citenamefont {Yeyati}, \citenamefont {Nyg{\aa}rd},\ and\ \citenamefont {Krogstrup}}]{Tosi:PRX19}%
  \BibitemOpen
  \bibfield  {author} {\bibinfo {author} {\bibfnamefont {L.}~\bibnamefont {Tosi}}, \bibinfo {author} {\bibfnamefont {C.}~\bibnamefont {Metzger}}, \bibinfo {author} {\bibfnamefont {M.~F.}\ \bibnamefont {Goffman}}, \bibinfo {author} {\bibfnamefont {C.}~\bibnamefont {Urbina}}, \bibinfo {author} {\bibfnamefont {H.}~\bibnamefont {Pothier}}, \bibinfo {author} {\bibfnamefont {S.}~\bibnamefont {Park}}, \bibinfo {author} {\bibfnamefont {A.~L.}\ \bibnamefont {Yeyati}}, \bibinfo {author} {\bibfnamefont {J.}~\bibnamefont {Nyg{\aa}rd}},\ and\ \bibinfo {author} {\bibfnamefont {P.}~\bibnamefont {Krogstrup}},\ }\bibfield  {title} {\bibinfo {title} {Spin-{{Orbit Splitting}} of {{Andreev States Revealed}} by {{Microwave Spectroscopy}}},\ }\href {https://doi.org/10.1103/PhysRevX.9.011010} {\bibfield  {journal} {\bibinfo  {journal} {Phys. Rev. X}\ }\textbf {\bibinfo {volume} {9}},\ \bibinfo {pages} {011010} (\bibinfo {year} {2019})}\BibitemShut {NoStop}%
\bibitem [{\citenamefont {{Matute-Ca{\~n}adas}}\ \emph {et~al.}(2022)\citenamefont {{Matute-Ca{\~n}adas}}, \citenamefont {Metzger}, \citenamefont {Park}, \citenamefont {Tosi}, \citenamefont {Krogstrup}, \citenamefont {Nyg{\aa}rd}, \citenamefont {Goffman}, \citenamefont {Urbina}, \citenamefont {Pothier},\ and\ \citenamefont {Yeyati}}]{Matute-Canadas:PRL22}%
  \BibitemOpen
  \bibfield  {author} {\bibinfo {author} {\bibfnamefont {F.~J.}\ \bibnamefont {{Matute-Ca{\~n}adas}}}, \bibinfo {author} {\bibfnamefont {C.}~\bibnamefont {Metzger}}, \bibinfo {author} {\bibfnamefont {S.}~\bibnamefont {Park}}, \bibinfo {author} {\bibfnamefont {L.}~\bibnamefont {Tosi}}, \bibinfo {author} {\bibfnamefont {P.}~\bibnamefont {Krogstrup}}, \bibinfo {author} {\bibfnamefont {J.}~\bibnamefont {Nyg{\aa}rd}}, \bibinfo {author} {\bibfnamefont {M.~F.}\ \bibnamefont {Goffman}}, \bibinfo {author} {\bibfnamefont {C.}~\bibnamefont {Urbina}}, \bibinfo {author} {\bibfnamefont {H.}~\bibnamefont {Pothier}},\ and\ \bibinfo {author} {\bibfnamefont {A.~L.}\ \bibnamefont {Yeyati}},\ }\bibfield  {title} {\bibinfo {title} {Signatures of {{Interactions}} in the {{Andreev Spectrum}} of {{Nanowire Josephson Junctions}}},\ }\href {https://doi.org/10.1103/PhysRevLett.128.197702} {\bibfield  {journal} {\bibinfo  {journal} {Phys. Rev. Lett.}\ }\textbf {\bibinfo {volume} {128}},\ \bibinfo {pages} {197702} (\bibinfo {year}
  {2022})}\BibitemShut {NoStop}%
\bibitem [{\citenamefont {Giavaras}\ and\ \citenamefont {Aguado}(2024)}]{Giavaras:PRB24}%
  \BibitemOpen
  \bibfield  {author} {\bibinfo {author} {\bibfnamefont {G.}~\bibnamefont {Giavaras}}\ and\ \bibinfo {author} {\bibfnamefont {R.}~\bibnamefont {Aguado}},\ }\bibfield  {title} {\bibinfo {title} {Flux-tunable supercurrent in full-shell nanowire {{Josephson}} junctions},\ }\href {https://doi.org/10.1103/PhysRevB.109.024509} {\bibfield  {journal} {\bibinfo  {journal} {Phys. Rev. B}\ }\textbf {\bibinfo {volume} {109}},\ \bibinfo {pages} {024509} (\bibinfo {year} {2024})}\BibitemShut {NoStop}%
\bibitem [{\citenamefont {Pay{\'a}}\ \emph {et~al.}(2025)\citenamefont {Pay{\'a}}, \citenamefont {Aguado}, \citenamefont {{San-Jose}},\ and\ \citenamefont {Prada}}]{Paya:PRB25}%
  \BibitemOpen
  \bibfield  {author} {\bibinfo {author} {\bibfnamefont {C.}~\bibnamefont {Pay{\'a}}}, \bibinfo {author} {\bibfnamefont {R.}~\bibnamefont {Aguado}}, \bibinfo {author} {\bibfnamefont {P.}~\bibnamefont {{San-Jose}}},\ and\ \bibinfo {author} {\bibfnamefont {E.}~\bibnamefont {Prada}},\ }\bibfield  {title} {\bibinfo {title} {Josephson effect and critical currents in trivial and topological full-shell hybrid nanowires},\ }\href {https://doi.org/10.1103/8mzs-dx7h} {\bibfield  {journal} {\bibinfo  {journal} {Phys. Rev. B}\ }\textbf {\bibinfo {volume} {111}},\ \bibinfo {pages} {235420} (\bibinfo {year} {2025})}\BibitemShut {NoStop}%
\bibitem [{\citenamefont {Woods}\ \emph {et~al.}(2019)\citenamefont {Woods}, \citenamefont {Das~Sarma},\ and\ \citenamefont {Stanescu}}]{Woods:PRB19}%
  \BibitemOpen
  \bibfield  {author} {\bibinfo {author} {\bibfnamefont {B.~D.}\ \bibnamefont {Woods}}, \bibinfo {author} {\bibfnamefont {S.}~\bibnamefont {Das~Sarma}},\ and\ \bibinfo {author} {\bibfnamefont {T.~D.}\ \bibnamefont {Stanescu}},\ }\bibfield  {title} {\bibinfo {title} {Electronic structure of full-shell {{InAs}}/{{Al}} hybrid semiconductor-superconductor nanowires: {{Spin-orbit}} coupling and topological phase space},\ }\href {https://doi.org/10.1103/PhysRevB.99.161118} {\bibfield  {journal} {\bibinfo  {journal} {Phys. Rev. B}\ }\textbf {\bibinfo {volume} {99}},\ \bibinfo {pages} {161118} (\bibinfo {year} {2019})}\BibitemShut {NoStop}%
\bibitem [{\citenamefont {Pe{\~n}aranda}\ \emph {et~al.}(2020)\citenamefont {Pe{\~n}aranda}, \citenamefont {Aguado}, \citenamefont {{San-Jose}},\ and\ \citenamefont {Prada}}]{Penaranda:PRR20}%
  \BibitemOpen
  \bibfield  {author} {\bibinfo {author} {\bibfnamefont {F.}~\bibnamefont {Pe{\~n}aranda}}, \bibinfo {author} {\bibfnamefont {R.}~\bibnamefont {Aguado}}, \bibinfo {author} {\bibfnamefont {P.}~\bibnamefont {{San-Jose}}},\ and\ \bibinfo {author} {\bibfnamefont {E.}~\bibnamefont {Prada}},\ }\bibfield  {title} {\bibinfo {title} {Even-odd effect and {{Majorana}} states in full-shell nanowires},\ }\href {https://doi.org/10.1103/PhysRevResearch.2.023171} {\bibfield  {journal} {\bibinfo  {journal} {Phys. Rev. Res.}\ }\textbf {\bibinfo {volume} {2}},\ \bibinfo {pages} {023171} (\bibinfo {year} {2020})}\BibitemShut {NoStop}%
\bibitem [{\citenamefont {Kopasov}\ and\ \citenamefont {Mel'nikov}(2020{\natexlab{a}})}]{Kopasov:PSS20}%
  \BibitemOpen
  \bibfield  {author} {\bibinfo {author} {\bibfnamefont {A.~A.}\ \bibnamefont {Kopasov}}\ and\ \bibinfo {author} {\bibfnamefont {A.~S.}\ \bibnamefont {Mel'nikov}},\ }\bibfield  {title} {\bibinfo {title} {Influence of the {{Accumulation Layer}} on the {{Spectral Properties}} of {{Full-Shell Majorana Nanowires}}},\ }\href {https://doi.org/10.1134/S1063783420090164} {\bibfield  {journal} {\bibinfo  {journal} {Phys. Solid State}\ }\textbf {\bibinfo {volume} {62}},\ \bibinfo {pages} {1592} (\bibinfo {year} {2020}{\natexlab{a}})}\BibitemShut {NoStop}%
\bibitem [{\citenamefont {Kopasov}\ and\ \citenamefont {Mel'nikov}(2020{\natexlab{b}})}]{Kopasov:PRB20}%
  \BibitemOpen
  \bibfield  {author} {\bibinfo {author} {\bibfnamefont {A.~A.}\ \bibnamefont {Kopasov}}\ and\ \bibinfo {author} {\bibfnamefont {A.~S.}\ \bibnamefont {Mel'nikov}},\ }\bibfield  {title} {\bibinfo {title} {Multiple topological transitions driven by the interplay of normal scattering and {{Andreev}} scattering},\ }\href {https://doi.org/10.1103/PhysRevB.101.054515} {\bibfield  {journal} {\bibinfo  {journal} {Phys. Rev. B}\ }\textbf {\bibinfo {volume} {101}},\ \bibinfo {pages} {054515} (\bibinfo {year} {2020}{\natexlab{b}})}\BibitemShut {NoStop}%
\bibitem [{\citenamefont {Metzger}\ \emph {et~al.}(2021)\citenamefont {Metzger}, \citenamefont {Park}, \citenamefont {Tosi}, \citenamefont {Janvier}, \citenamefont {Reynoso}, \citenamefont {Goffman}, \citenamefont {Urbina}, \citenamefont {Levy~Yeyati},\ and\ \citenamefont {Pothier}}]{Metzger:PRR21}%
  \BibitemOpen
  \bibfield  {author} {\bibinfo {author} {\bibfnamefont {C.}~\bibnamefont {Metzger}}, \bibinfo {author} {\bibfnamefont {S.}~\bibnamefont {Park}}, \bibinfo {author} {\bibfnamefont {L.}~\bibnamefont {Tosi}}, \bibinfo {author} {\bibfnamefont {C.}~\bibnamefont {Janvier}}, \bibinfo {author} {\bibfnamefont {A.~A.}\ \bibnamefont {Reynoso}}, \bibinfo {author} {\bibfnamefont {M.~F.}\ \bibnamefont {Goffman}}, \bibinfo {author} {\bibfnamefont {C.}~\bibnamefont {Urbina}}, \bibinfo {author} {\bibfnamefont {A.}~\bibnamefont {Levy~Yeyati}},\ and\ \bibinfo {author} {\bibfnamefont {H.}~\bibnamefont {Pothier}},\ }\bibfield  {title} {\bibinfo {title} {Circuit-{{QED}} with phase-biased {{Josephson}} weak links},\ }\href {https://doi.org/10.1103/PhysRevResearch.3.013036} {\bibfield  {journal} {\bibinfo  {journal} {Phys. Rev. Res.}\ }\textbf {\bibinfo {volume} {3}},\ \bibinfo {pages} {013036} (\bibinfo {year} {2021})}\BibitemShut {NoStop}%
\bibitem [{\citenamefont {Escribano}\ \emph {et~al.}(2022)\citenamefont {Escribano}, \citenamefont {Levy~Yeyati}, \citenamefont {Aguado}, \citenamefont {Prada},\ and\ \citenamefont {{San-Jose}}}]{Escribano:PRB22}%
  \BibitemOpen
  \bibfield  {author} {\bibinfo {author} {\bibfnamefont {S.~D.}\ \bibnamefont {Escribano}}, \bibinfo {author} {\bibfnamefont {A.}~\bibnamefont {Levy~Yeyati}}, \bibinfo {author} {\bibfnamefont {R.}~\bibnamefont {Aguado}}, \bibinfo {author} {\bibfnamefont {E.}~\bibnamefont {Prada}},\ and\ \bibinfo {author} {\bibfnamefont {P.}~\bibnamefont {{San-Jose}}},\ }\bibfield  {title} {\bibinfo {title} {Fluxoid-induced pairing suppression and near-zero modes in quantum dots coupled to full-shell nanowires},\ }\href {https://doi.org/10.1103/PhysRevB.105.045418} {\bibfield  {journal} {\bibinfo  {journal} {Phys. Rev. B}\ }\textbf {\bibinfo {volume} {105}},\ \bibinfo {pages} {045418} (\bibinfo {year} {2022})}\BibitemShut {NoStop}%
\bibitem [{\citenamefont {{San-Jose}}\ \emph {et~al.}(2023)\citenamefont {{San-Jose}}, \citenamefont {Pay{\'a}}, \citenamefont {Marcus}, \citenamefont {Vaitiek{\.e}nas},\ and\ \citenamefont {Prada}}]{San-Jose:PRB23}%
  \BibitemOpen
  \bibfield  {author} {\bibinfo {author} {\bibfnamefont {P.}~\bibnamefont {{San-Jose}}}, \bibinfo {author} {\bibfnamefont {C.}~\bibnamefont {Pay{\'a}}}, \bibinfo {author} {\bibfnamefont {C.~M.}\ \bibnamefont {Marcus}}, \bibinfo {author} {\bibfnamefont {S.}~\bibnamefont {Vaitiek{\.e}nas}},\ and\ \bibinfo {author} {\bibfnamefont {E.}~\bibnamefont {Prada}},\ }\bibfield  {title} {\bibinfo {title} {Theory of {{Caroli--de Gennes--Matricon}} analogs in full-shell hybrid nanowires},\ }\href {https://doi.org/10.1103/PhysRevB.107.155423} {\bibfield  {journal} {\bibinfo  {journal} {Phys. Rev. B}\ }\textbf {\bibinfo {volume} {107}},\ \bibinfo {pages} {155423} (\bibinfo {year} {2023})}\BibitemShut {NoStop}%
\bibitem [{\citenamefont {Pay{\'a}}\ \emph {et~al.}(2024{\natexlab{a}})\citenamefont {Pay{\'a}}, \citenamefont {Escribano}, \citenamefont {Vezzosi}, \citenamefont {Pe{\~n}aranda}, \citenamefont {Aguado}, \citenamefont {{San-Jose}},\ and\ \citenamefont {Prada}}]{Paya:PRB24}%
  \BibitemOpen
  \bibfield  {author} {\bibinfo {author} {\bibfnamefont {C.}~\bibnamefont {Pay{\'a}}}, \bibinfo {author} {\bibfnamefont {S.~D.}\ \bibnamefont {Escribano}}, \bibinfo {author} {\bibfnamefont {A.}~\bibnamefont {Vezzosi}}, \bibinfo {author} {\bibfnamefont {F.}~\bibnamefont {Pe{\~n}aranda}}, \bibinfo {author} {\bibfnamefont {R.}~\bibnamefont {Aguado}}, \bibinfo {author} {\bibfnamefont {P.}~\bibnamefont {{San-Jose}}},\ and\ \bibinfo {author} {\bibfnamefont {E.}~\bibnamefont {Prada}},\ }\bibfield  {title} {\bibinfo {title} {Phenomenology of {{Majorana}} zero modes in full-shell hybrid nanowires},\ }\href {https://doi.org/10.1103/PhysRevB.109.115428} {\bibfield  {journal} {\bibinfo  {journal} {Phys. Rev. B}\ }\textbf {\bibinfo {volume} {109}},\ \bibinfo {pages} {115428} (\bibinfo {year} {2024}{\natexlab{a}})}\BibitemShut {NoStop}%
\bibitem [{\citenamefont {Pay{\'a}}\ \emph {et~al.}(2024{\natexlab{b}})\citenamefont {Pay{\'a}}, \citenamefont {{San-Jose}}, \citenamefont {Mart{\'i}nez}, \citenamefont {Aguado},\ and\ \citenamefont {Prada}}]{Paya:PRB24a}%
  \BibitemOpen
  \bibfield  {author} {\bibinfo {author} {\bibfnamefont {C.}~\bibnamefont {Pay{\'a}}}, \bibinfo {author} {\bibfnamefont {P.}~\bibnamefont {{San-Jose}}}, \bibinfo {author} {\bibfnamefont {C.~J.~S.}\ \bibnamefont {Mart{\'i}nez}}, \bibinfo {author} {\bibfnamefont {R.}~\bibnamefont {Aguado}},\ and\ \bibinfo {author} {\bibfnamefont {E.}~\bibnamefont {Prada}},\ }\bibfield  {title} {\bibinfo {title} {Absence of {{Majorana}} oscillations in finite-length full-shell hybrid nanowires},\ }\href {https://doi.org/10.1103/PhysRevB.110.115417} {\bibfield  {journal} {\bibinfo  {journal} {Phys. Rev. B}\ }\textbf {\bibinfo {volume} {110}},\ \bibinfo {pages} {115417} (\bibinfo {year} {2024}{\natexlab{b}})}\BibitemShut {NoStop}%
\bibitem [{\citenamefont {Vezzosi}\ \emph {et~al.}(2025)\citenamefont {Vezzosi}, \citenamefont {Pay{\'a}}, \citenamefont {W{\'o}jcik}, \citenamefont {Bertoni}, \citenamefont {Goldoni}, \citenamefont {Prada},\ and\ \citenamefont {D.~Escribano}}]{Vezzosi:SP25}%
  \BibitemOpen
  \bibfield  {author} {\bibinfo {author} {\bibfnamefont {A.}~\bibnamefont {Vezzosi}}, \bibinfo {author} {\bibfnamefont {C.}~\bibnamefont {Pay{\'a}}}, \bibinfo {author} {\bibfnamefont {P.}~\bibnamefont {W{\'o}jcik}}, \bibinfo {author} {\bibfnamefont {A.}~\bibnamefont {Bertoni}}, \bibinfo {author} {\bibfnamefont {G.}~\bibnamefont {Goldoni}}, \bibinfo {author} {\bibfnamefont {E.}~\bibnamefont {Prada}},\ and\ \bibinfo {author} {\bibfnamefont {S.}~\bibnamefont {D.~Escribano}},\ }\bibfield  {title} {\bibinfo {title} {{{InP}}/{{GaSb}} core-shell nanowires: {{A}} novel hole-based platform with strong spin-orbit coupling for full-shell hybrid devices},\ }\href {https://doi.org/10.21468/SciPostPhys.18.2.069} {\bibfield  {journal} {\bibinfo  {journal} {SciPost Physics}\ }\textbf {\bibinfo {volume} {18}},\ \bibinfo {pages} {069} (\bibinfo {year} {2025})}\BibitemShut {NoStop}%
\bibitem [{\citenamefont {Vaitiek{\.e}nas}\ \emph {et~al.}(2020{\natexlab{a}})\citenamefont {Vaitiek{\.e}nas}, \citenamefont {Krogstrup},\ and\ \citenamefont {Marcus}}]{Vaitiekenas:PRB20}%
  \BibitemOpen
  \bibfield  {author} {\bibinfo {author} {\bibfnamefont {S.}~\bibnamefont {Vaitiek{\.e}nas}}, \bibinfo {author} {\bibfnamefont {P.}~\bibnamefont {Krogstrup}},\ and\ \bibinfo {author} {\bibfnamefont {C.~M.}\ \bibnamefont {Marcus}},\ }\bibfield  {title} {\bibinfo {title} {Anomalous metallic phase in tunable destructive superconductors},\ }\href {https://doi.org/10.1103/PhysRevB.101.060507} {\bibfield  {journal} {\bibinfo  {journal} {Phys. Rev. B}\ }\textbf {\bibinfo {volume} {101}},\ \bibinfo {pages} {060507} (\bibinfo {year} {2020}{\natexlab{a}})}\BibitemShut {NoStop}%
\bibitem [{\citenamefont {Vaitiek{\.e}nas}\ \emph {et~al.}(2020{\natexlab{b}})\citenamefont {Vaitiek{\.e}nas}, \citenamefont {Winkler}, \citenamefont {{van Heck}}, \citenamefont {Karzig}, \citenamefont {Deng}, \citenamefont {Flensberg}, \citenamefont {Glazman}, \citenamefont {Nayak}, \citenamefont {Krogstrup}, \citenamefont {Lutchyn},\ and\ \citenamefont {Marcus}}]{Vaitiekenas:S20}%
  \BibitemOpen
  \bibfield  {author} {\bibinfo {author} {\bibfnamefont {S.}~\bibnamefont {Vaitiek{\.e}nas}}, \bibinfo {author} {\bibfnamefont {G.~W.}\ \bibnamefont {Winkler}}, \bibinfo {author} {\bibfnamefont {B.}~\bibnamefont {{van Heck}}}, \bibinfo {author} {\bibfnamefont {T.}~\bibnamefont {Karzig}}, \bibinfo {author} {\bibfnamefont {M.-T.}\ \bibnamefont {Deng}}, \bibinfo {author} {\bibfnamefont {K.}~\bibnamefont {Flensberg}}, \bibinfo {author} {\bibfnamefont {L.~I.}\ \bibnamefont {Glazman}}, \bibinfo {author} {\bibfnamefont {C.}~\bibnamefont {Nayak}}, \bibinfo {author} {\bibfnamefont {P.}~\bibnamefont {Krogstrup}}, \bibinfo {author} {\bibfnamefont {R.~M.}\ \bibnamefont {Lutchyn}},\ and\ \bibinfo {author} {\bibfnamefont {C.~M.}\ \bibnamefont {Marcus}},\ }\bibfield  {title} {\bibinfo {title} {Flux-induced topological superconductivity in full-shell nanowires},\ }\href {https://doi.org/10.1126/science.aav3392} {\bibfield  {journal} {\bibinfo  {journal} {Science}\ }\textbf {\bibinfo {volume} {367}},\ \bibinfo {pages}
  {eaav3392} (\bibinfo {year} {2020}{\natexlab{b}})}\BibitemShut {NoStop}%
\bibitem [{Vai(2025)}]{Vaitiekenas:ErrS25}%
  \BibitemOpen
  \bibfield  {title} {\bibinfo {title} {Erratum for the {{Research Article}} ``{{Flux-induced}} topological superconductivity in full-shell nanowires'' by {{S}}. {{Vaitiek{\.e}nas}} et al.},\ }\href {https://doi.org/10.1126/science.aea6837} {\bibfield  {journal} {\bibinfo  {journal} {Science}\ }\textbf {\bibinfo {volume} {389}},\ \bibinfo {pages} {eaea6837} (\bibinfo {year} {2025})}\BibitemShut {NoStop}%
\bibitem [{\citenamefont {Valentini}\ \emph {et~al.}(2021)\citenamefont {Valentini}, \citenamefont {Pe{\~n}aranda}, \citenamefont {Hofmann}, \citenamefont {Brauns}, \citenamefont {Hauschild}, \citenamefont {Krogstrup}, \citenamefont {{San-Jose}}, \citenamefont {Prada}, \citenamefont {Aguado},\ and\ \citenamefont {Katsaros}}]{Valentini:S21}%
  \BibitemOpen
  \bibfield  {author} {\bibinfo {author} {\bibfnamefont {M.}~\bibnamefont {Valentini}}, \bibinfo {author} {\bibfnamefont {F.}~\bibnamefont {Pe{\~n}aranda}}, \bibinfo {author} {\bibfnamefont {A.}~\bibnamefont {Hofmann}}, \bibinfo {author} {\bibfnamefont {M.}~\bibnamefont {Brauns}}, \bibinfo {author} {\bibfnamefont {R.}~\bibnamefont {Hauschild}}, \bibinfo {author} {\bibfnamefont {P.}~\bibnamefont {Krogstrup}}, \bibinfo {author} {\bibfnamefont {P.}~\bibnamefont {{San-Jose}}}, \bibinfo {author} {\bibfnamefont {E.}~\bibnamefont {Prada}}, \bibinfo {author} {\bibfnamefont {R.}~\bibnamefont {Aguado}},\ and\ \bibinfo {author} {\bibfnamefont {G.}~\bibnamefont {Katsaros}},\ }\bibfield  {title} {\bibinfo {title} {Nontopological zero-bias peaks in full-shell nanowires induced by flux-tunable {{Andreev}} states},\ }\href {https://doi.org/10.1126/science.abf1513} {\bibfield  {journal} {\bibinfo  {journal} {Science}\ }\textbf {\bibinfo {volume} {373}},\ \bibinfo {pages} {82} (\bibinfo {year} {2021})}\BibitemShut {NoStop}%
\bibitem [{\citenamefont {Valentini}\ \emph {et~al.}(2022)\citenamefont {Valentini}, \citenamefont {Borovkov}, \citenamefont {Prada}, \citenamefont {{Mart{\'i}-S{\'a}nchez}}, \citenamefont {Botifoll}, \citenamefont {Hofmann}, \citenamefont {Arbiol}, \citenamefont {Aguado}, \citenamefont {{San-Jose}},\ and\ \citenamefont {Katsaros}}]{Valentini:N22}%
  \BibitemOpen
  \bibfield  {author} {\bibinfo {author} {\bibfnamefont {M.}~\bibnamefont {Valentini}}, \bibinfo {author} {\bibfnamefont {M.}~\bibnamefont {Borovkov}}, \bibinfo {author} {\bibfnamefont {E.}~\bibnamefont {Prada}}, \bibinfo {author} {\bibfnamefont {S.}~\bibnamefont {{Mart{\'i}-S{\'a}nchez}}}, \bibinfo {author} {\bibfnamefont {M.}~\bibnamefont {Botifoll}}, \bibinfo {author} {\bibfnamefont {A.}~\bibnamefont {Hofmann}}, \bibinfo {author} {\bibfnamefont {J.}~\bibnamefont {Arbiol}}, \bibinfo {author} {\bibfnamefont {R.}~\bibnamefont {Aguado}}, \bibinfo {author} {\bibfnamefont {P.}~\bibnamefont {{San-Jose}}},\ and\ \bibinfo {author} {\bibfnamefont {G.}~\bibnamefont {Katsaros}},\ }\bibfield  {title} {\bibinfo {title} {Majorana-like {{Coulomb}} spectroscopy in the absence of zero-bias peaks},\ }\href {https://doi.org/10.1038/s41586-022-05382-w} {\bibfield  {journal} {\bibinfo  {journal} {Nature}\ }\textbf {\bibinfo {volume} {612}},\ \bibinfo {pages} {442} (\bibinfo {year} {2022})}\BibitemShut {NoStop}%
\bibitem [{\citenamefont {Valentini}\ \emph {et~al.}(2025)\citenamefont {Valentini}, \citenamefont {Souto}, \citenamefont {Borovkov}, \citenamefont {Krogstrup}, \citenamefont {Meir}, \citenamefont {Leijnse}, \citenamefont {Danon},\ and\ \citenamefont {Katsaros}}]{Valentini:PRR25}%
  \BibitemOpen
  \bibfield  {author} {\bibinfo {author} {\bibfnamefont {M.}~\bibnamefont {Valentini}}, \bibinfo {author} {\bibfnamefont {R.~S.}\ \bibnamefont {Souto}}, \bibinfo {author} {\bibfnamefont {M.}~\bibnamefont {Borovkov}}, \bibinfo {author} {\bibfnamefont {P.}~\bibnamefont {Krogstrup}}, \bibinfo {author} {\bibfnamefont {Y.}~\bibnamefont {Meir}}, \bibinfo {author} {\bibfnamefont {M.}~\bibnamefont {Leijnse}}, \bibinfo {author} {\bibfnamefont {J.}~\bibnamefont {Danon}},\ and\ \bibinfo {author} {\bibfnamefont {G.}~\bibnamefont {Katsaros}},\ }\bibfield  {title} {\bibinfo {title} {Subgap transport in superconductor-semiconductor hybrid islands: {{Weak}} and strong coupling regimes},\ }\href {https://doi.org/10.1103/PhysRevResearch.7.023022} {\bibfield  {journal} {\bibinfo  {journal} {Phys. Rev. Res.}\ }\textbf {\bibinfo {volume} {7}},\ \bibinfo {pages} {023022} (\bibinfo {year} {2025})}\BibitemShut {NoStop}%
\bibitem [{\citenamefont {Deng}\ \emph {et~al.}(2025)\citenamefont {Deng}, \citenamefont {Pay{\'a}}, \citenamefont {{San-Jose}}, \citenamefont {Prada}, \citenamefont {Marcus},\ and\ \citenamefont {Vaitiek{\.e}nas}}]{Deng:PRL25}%
  \BibitemOpen
  \bibfield  {author} {\bibinfo {author} {\bibfnamefont {M.~T.}\ \bibnamefont {Deng}}, \bibinfo {author} {\bibfnamefont {C.}~\bibnamefont {Pay{\'a}}}, \bibinfo {author} {\bibfnamefont {P.}~\bibnamefont {{San-Jose}}}, \bibinfo {author} {\bibfnamefont {E.}~\bibnamefont {Prada}}, \bibinfo {author} {\bibfnamefont {C.~M.}\ \bibnamefont {Marcus}},\ and\ \bibinfo {author} {\bibfnamefont {S.}~\bibnamefont {Vaitiek{\.e}nas}},\ }\bibfield  {title} {\bibinfo {title} {Caroli--de {{Gennes--Matricon Analogs}} in {{Full-Shell Hybrid Nanowires}}},\ }\href {https://doi.org/10.1103/PhysRevLett.134.206302} {\bibfield  {journal} {\bibinfo  {journal} {Phys. Rev. Lett.}\ }\textbf {\bibinfo {volume} {134}},\ \bibinfo {pages} {206302} (\bibinfo {year} {2025})}\BibitemShut {NoStop}%
\bibitem [{\citenamefont {Little}\ and\ \citenamefont {Parks}(1962)}]{Little:PRL62}%
  \BibitemOpen
  \bibfield  {author} {\bibinfo {author} {\bibfnamefont {W.~A.}\ \bibnamefont {Little}}\ and\ \bibinfo {author} {\bibfnamefont {R.~D.}\ \bibnamefont {Parks}},\ }\bibfield  {title} {\bibinfo {title} {Observation of {{Quantum Periodicity}} in the {{Transition Temperature}} of a {{Superconducting Cylinder}}},\ }\href {https://doi.org/10.1103/PhysRevLett.9.9} {\bibfield  {journal} {\bibinfo  {journal} {Phys. Rev. Lett.}\ }\textbf {\bibinfo {volume} {9}},\ \bibinfo {pages} {9} (\bibinfo {year} {1962})}\BibitemShut {NoStop}%
\bibitem [{\citenamefont {Parks}\ and\ \citenamefont {Little}(1964)}]{Parks:PR64}%
  \BibitemOpen
  \bibfield  {author} {\bibinfo {author} {\bibfnamefont {R.~D.}\ \bibnamefont {Parks}}\ and\ \bibinfo {author} {\bibfnamefont {W.~A.}\ \bibnamefont {Little}},\ }\bibfield  {title} {\bibinfo {title} {Fluxoid {{Quantization}} in a {{Multiply-Connected Superconductor}}},\ }\href {https://doi.org/10.1103/PhysRev.133.A97} {\bibfield  {journal} {\bibinfo  {journal} {Phys. Rev.}\ }\textbf {\bibinfo {volume} {133}},\ \bibinfo {pages} {A97} (\bibinfo {year} {1964})}\BibitemShut {NoStop}%
\bibitem [{\citenamefont {Tinkham}(1996)}]{Tinkham:96}%
  \BibitemOpen
  \bibfield  {author} {\bibinfo {author} {\bibfnamefont {M.}~\bibnamefont {Tinkham}},\ }\href@noop {} {\emph {\bibinfo {title} {Introduction to Superconductivity}}},\ \bibinfo {edition} {2nd}\ ed.,\ International Series in Pure and Applied Physics\ (\bibinfo  {publisher} {McGraw Hill},\ \bibinfo {address} {New York},\ \bibinfo {year} {1996})\BibitemShut {NoStop}%
\bibitem [{\citenamefont {Qi}\ and\ \citenamefont {Zhang}(2011)}]{Qi:RMP11}%
  \BibitemOpen
  \bibfield  {author} {\bibinfo {author} {\bibfnamefont {X.-L.}\ \bibnamefont {Qi}}\ and\ \bibinfo {author} {\bibfnamefont {S.-C.}\ \bibnamefont {Zhang}},\ }\bibfield  {title} {\bibinfo {title} {Topological insulators and superconductors},\ }\href {https://doi.org/10.1103/RevModPhys.83.1057} {\bibfield  {journal} {\bibinfo  {journal} {Rev. Mod. Phys.}\ }\textbf {\bibinfo {volume} {83}},\ \bibinfo {pages} {1057} (\bibinfo {year} {2011})}\BibitemShut {NoStop}%
\bibitem [{\citenamefont {Alicea}(2012)}]{Alicea:RPP12}%
  \BibitemOpen
  \bibfield  {author} {\bibinfo {author} {\bibfnamefont {J.}~\bibnamefont {Alicea}},\ }\bibfield  {title} {\bibinfo {title} {New directions in the pursuit of {{Majorana}} fermions in solid state systems},\ }\href {https://doi.org/10.1088/0034-4885/75/7/076501} {\bibfield  {journal} {\bibinfo  {journal} {Rep. Prog. Phys.}\ }\textbf {\bibinfo {volume} {75}},\ \bibinfo {pages} {076501} (\bibinfo {year} {2012})}\BibitemShut {NoStop}%
\bibitem [{\citenamefont {Aguado}(2017)}]{Aguado:RNC17}%
  \BibitemOpen
  \bibfield  {author} {\bibinfo {author} {\bibfnamefont {R.}~\bibnamefont {Aguado}},\ }\bibfield  {title} {\bibinfo {title} {Majorana quasiparticles in condensed matter},\ }\href {https://doi.org/10.1393/ncr/i2017-10141-9} {\bibfield  {journal} {\bibinfo  {journal} {Riv. Nuovo Cim.}\ }\textbf {\bibinfo {volume} {40}},\ \bibinfo {pages} {523} (\bibinfo {year} {2017})}\BibitemShut {NoStop}%
\bibitem [{\citenamefont {Prada}\ \emph {et~al.}(2020)\citenamefont {Prada}, \citenamefont {{San-Jose}}, \citenamefont {{de Moor}}, \citenamefont {Geresdi}, \citenamefont {Lee}, \citenamefont {Klinovaja}, \citenamefont {Loss}, \citenamefont {Nyg{\aa}rd}, \citenamefont {Aguado},\ and\ \citenamefont {Kouwenhoven}}]{Prada:NRP20}%
  \BibitemOpen
  \bibfield  {author} {\bibinfo {author} {\bibfnamefont {E.}~\bibnamefont {Prada}}, \bibinfo {author} {\bibfnamefont {P.}~\bibnamefont {{San-Jose}}}, \bibinfo {author} {\bibfnamefont {M.~W.~A.}\ \bibnamefont {{de Moor}}}, \bibinfo {author} {\bibfnamefont {A.}~\bibnamefont {Geresdi}}, \bibinfo {author} {\bibfnamefont {E.~J.~H.}\ \bibnamefont {Lee}}, \bibinfo {author} {\bibfnamefont {J.}~\bibnamefont {Klinovaja}}, \bibinfo {author} {\bibfnamefont {D.}~\bibnamefont {Loss}}, \bibinfo {author} {\bibfnamefont {J.}~\bibnamefont {Nyg{\aa}rd}}, \bibinfo {author} {\bibfnamefont {R.}~\bibnamefont {Aguado}},\ and\ \bibinfo {author} {\bibfnamefont {L.~P.}\ \bibnamefont {Kouwenhoven}},\ }\bibfield  {title} {\bibinfo {title} {From {{Andreev}} to {{Majorana}} bound states in hybrid superconductor--semiconductor nanowires},\ }\href {https://doi.org/10.1038/s42254-020-0228-y} {\bibfield  {journal} {\bibinfo  {journal} {Nat Rev Phys}\ }\textbf {\bibinfo {volume} {2}},\ \bibinfo {pages} {575} (\bibinfo {year}
  {2020})}\BibitemShut {NoStop}%
\bibitem [{\citenamefont {Vekris}\ \emph {et~al.}(2021)\citenamefont {Vekris}, \citenamefont {Estrada~Salda{\~n}a}, \citenamefont {{de Bruijckere}}, \citenamefont {Lori{\'c}}, \citenamefont {Kanne}, \citenamefont {Marnauza}, \citenamefont {Olsteins}, \citenamefont {Nyg{\aa}rd},\ and\ \citenamefont {{Grove-Rasmussen}}}]{Vekris:SR21}%
  \BibitemOpen
  \bibfield  {author} {\bibinfo {author} {\bibfnamefont {A.}~\bibnamefont {Vekris}}, \bibinfo {author} {\bibfnamefont {J.~C.}\ \bibnamefont {Estrada~Salda{\~n}a}}, \bibinfo {author} {\bibfnamefont {J.}~\bibnamefont {{de Bruijckere}}}, \bibinfo {author} {\bibfnamefont {S.}~\bibnamefont {Lori{\'c}}}, \bibinfo {author} {\bibfnamefont {T.}~\bibnamefont {Kanne}}, \bibinfo {author} {\bibfnamefont {M.}~\bibnamefont {Marnauza}}, \bibinfo {author} {\bibfnamefont {D.}~\bibnamefont {Olsteins}}, \bibinfo {author} {\bibfnamefont {J.}~\bibnamefont {Nyg{\aa}rd}},\ and\ \bibinfo {author} {\bibfnamefont {K.}~\bibnamefont {{Grove-Rasmussen}}},\ }\bibfield  {title} {\bibinfo {title} {Asymmetric {{Little}}--{{Parks}} oscillations in full shell double nanowires},\ }\href {https://doi.org/10.1038/s41598-021-97780-9} {\bibfield  {journal} {\bibinfo  {journal} {Sci Rep}\ }\textbf {\bibinfo {volume} {11}},\ \bibinfo {pages} {19034} (\bibinfo {year} {2021})}\BibitemShut {NoStop}%
\bibitem [{\citenamefont {Ibabe}\ \emph {et~al.}(2023)\citenamefont {Ibabe}, \citenamefont {G{\'o}mez}, \citenamefont {Steffensen}, \citenamefont {Kanne}, \citenamefont {Nyg{\aa}rd}, \citenamefont {Yeyati},\ and\ \citenamefont {Lee}}]{Ibabe:NC23}%
  \BibitemOpen
  \bibfield  {author} {\bibinfo {author} {\bibfnamefont {A.}~\bibnamefont {Ibabe}}, \bibinfo {author} {\bibfnamefont {M.}~\bibnamefont {G{\'o}mez}}, \bibinfo {author} {\bibfnamefont {G.~O.}\ \bibnamefont {Steffensen}}, \bibinfo {author} {\bibfnamefont {T.}~\bibnamefont {Kanne}}, \bibinfo {author} {\bibfnamefont {J.}~\bibnamefont {Nyg{\aa}rd}}, \bibinfo {author} {\bibfnamefont {A.~L.}\ \bibnamefont {Yeyati}},\ and\ \bibinfo {author} {\bibfnamefont {E.~J.~H.}\ \bibnamefont {Lee}},\ }\bibfield  {title} {\bibinfo {title} {Joule spectroscopy of hybrid superconductor--semiconductor nanodevices},\ }\href {https://doi.org/10.1038/s41467-023-38533-2} {\bibfield  {journal} {\bibinfo  {journal} {Nat Commun}\ }\textbf {\bibinfo {volume} {14}},\ \bibinfo {pages} {2873} (\bibinfo {year} {2023})}\BibitemShut {NoStop}%
\bibitem [{\citenamefont {Ibabe}\ \emph {et~al.}(2024)\citenamefont {Ibabe}, \citenamefont {Steffensen}, \citenamefont {Casal}, \citenamefont {G{\'o}mez}, \citenamefont {Kanne}, \citenamefont {Nyg{\aa}rd}, \citenamefont {Levy~Yeyati},\ and\ \citenamefont {Lee}}]{Ibabe:NL24}%
  \BibitemOpen
  \bibfield  {author} {\bibinfo {author} {\bibfnamefont {{\'A}.}~\bibnamefont {Ibabe}}, \bibinfo {author} {\bibfnamefont {G.~O.}\ \bibnamefont {Steffensen}}, \bibinfo {author} {\bibfnamefont {I.}~\bibnamefont {Casal}}, \bibinfo {author} {\bibfnamefont {M.}~\bibnamefont {G{\'o}mez}}, \bibinfo {author} {\bibfnamefont {T.}~\bibnamefont {Kanne}}, \bibinfo {author} {\bibfnamefont {J.}~\bibnamefont {Nyg{\aa}rd}}, \bibinfo {author} {\bibfnamefont {A.}~\bibnamefont {Levy~Yeyati}},\ and\ \bibinfo {author} {\bibfnamefont {E.~J.~H.}\ \bibnamefont {Lee}},\ }\bibfield  {title} {\bibinfo {title} {Heat {{Dissipation Mechanisms}} in {{Hybrid Superconductor}}--{{Semiconductor Devices Revealed}} by {{Joule Spectroscopy}}},\ }\href {https://doi.org/10.1021/acs.nanolett.4c00574} {\bibfield  {journal} {\bibinfo  {journal} {Nano Lett.}\ }\textbf {\bibinfo {volume} {24}},\ \bibinfo {pages} {6488} (\bibinfo {year} {2024})}\BibitemShut {NoStop}%
\bibitem [{\citenamefont {Ercolani}\ \emph {et~al.}(2012)\citenamefont {Ercolani}, \citenamefont {Gemmi}, \citenamefont {Nasi}, \citenamefont {Rossi}, \citenamefont {Pea}, \citenamefont {Li}, \citenamefont {Salviati}, \citenamefont {Beltram},\ and\ \citenamefont {Sorba}}]{Ercolani:N12}%
  \BibitemOpen
  \bibfield  {author} {\bibinfo {author} {\bibfnamefont {D.}~\bibnamefont {Ercolani}}, \bibinfo {author} {\bibfnamefont {M.}~\bibnamefont {Gemmi}}, \bibinfo {author} {\bibfnamefont {L.}~\bibnamefont {Nasi}}, \bibinfo {author} {\bibfnamefont {F.}~\bibnamefont {Rossi}}, \bibinfo {author} {\bibfnamefont {M.}~\bibnamefont {Pea}}, \bibinfo {author} {\bibfnamefont {A.}~\bibnamefont {Li}}, \bibinfo {author} {\bibfnamefont {G.}~\bibnamefont {Salviati}}, \bibinfo {author} {\bibfnamefont {F.}~\bibnamefont {Beltram}},\ and\ \bibinfo {author} {\bibfnamefont {L.}~\bibnamefont {Sorba}},\ }\bibfield  {title} {\bibinfo {title} {Growth of {{InAs}}/{{InAsSb}} heterostructured nanowires},\ }\href {https://doi.org/10.1088/0957-4484/23/11/115606} {\bibfield  {journal} {\bibinfo  {journal} {Nanotechnology}\ }\textbf {\bibinfo {volume} {23}},\ \bibinfo {pages} {115606} (\bibinfo {year} {2012})}\BibitemShut {NoStop}%
\bibitem [{\citenamefont {Krogstrup}\ \emph {et~al.}(2015)\citenamefont {Krogstrup}, \citenamefont {Ziino}, \citenamefont {Chang}, \citenamefont {Albrecht}, \citenamefont {Madsen}, \citenamefont {Johnson}, \citenamefont {Nyg{\aa}rd}, \citenamefont {Marcus},\ and\ \citenamefont {Jespersen}}]{Krogstrup:NM15}%
  \BibitemOpen
  \bibfield  {author} {\bibinfo {author} {\bibfnamefont {P.}~\bibnamefont {Krogstrup}}, \bibinfo {author} {\bibfnamefont {N.~L.~B.}\ \bibnamefont {Ziino}}, \bibinfo {author} {\bibfnamefont {W.}~\bibnamefont {Chang}}, \bibinfo {author} {\bibfnamefont {S.~M.}\ \bibnamefont {Albrecht}}, \bibinfo {author} {\bibfnamefont {M.~H.}\ \bibnamefont {Madsen}}, \bibinfo {author} {\bibfnamefont {E.}~\bibnamefont {Johnson}}, \bibinfo {author} {\bibfnamefont {J.}~\bibnamefont {Nyg{\aa}rd}}, \bibinfo {author} {\bibfnamefont {C.~M.}\ \bibnamefont {Marcus}},\ and\ \bibinfo {author} {\bibfnamefont {T.~S.}\ \bibnamefont {Jespersen}},\ }\bibfield  {title} {\bibinfo {title} {Epitaxy of semiconductor--superconductor nanowires},\ }\href {https://doi.org/10.1038/nmat4176} {\bibfield  {journal} {\bibinfo  {journal} {Nature Mater}\ }\textbf {\bibinfo {volume} {14}},\ \bibinfo {pages} {400} (\bibinfo {year} {2015})}\BibitemShut {NoStop}%
\bibitem [{Note1()}]{Note1}%
  \BibitemOpen
  \bibinfo {note} {Note that the winding number $n(\Phi )$ is only defined in the $S_1$ and $S_2$ regions (where it takes integer values $n_1$ and $n_2$, respectively), but not in the $N$ region, since the latter does not have a superconducting order parameter. Hence, in the split-shell geometry of Fig. \ref {fig:sketch} considered here, the geometric details of the $N$ region do not affect the $n_1, n_2$ values.}\BibitemShut {Stop}%
\bibitem [{Note2()}]{Note2}%
  \BibitemOpen
  \bibinfo {note} {Note that topological superconductivity can also be found in even number LP lobes if the cylindrical symmetry of the hybrid wire is sufficiently broken~\cite {Vaitiekenas:S20,*Vaitiekenas:ErrS25,Paya:PRB24}.}\BibitemShut {Stop}%
\bibitem [{\citenamefont {Tilley}(1966)}]{Tilley:PL66}%
  \BibitemOpen
  \bibfield  {author} {\bibinfo {author} {\bibfnamefont {D.~R.}\ \bibnamefont {Tilley}},\ }\bibfield  {title} {\bibinfo {title} {Cylindrical {{Josephson}} junctions},\ }\href {https://doi.org/10.1016/0031-9163(66)90896-1} {\bibfield  {journal} {\bibinfo  {journal} {Physics Letters}\ }\textbf {\bibinfo {volume} {20}},\ \bibinfo {pages} {117} (\bibinfo {year} {1966})}\BibitemShut {NoStop}%
\bibitem [{\citenamefont {Sherrill}\ and\ \citenamefont {Bhushan}(1979)}]{Sherrill:PRB79}%
  \BibitemOpen
  \bibfield  {author} {\bibinfo {author} {\bibfnamefont {M.~D.}\ \bibnamefont {Sherrill}}\ and\ \bibinfo {author} {\bibfnamefont {M.}~\bibnamefont {Bhushan}},\ }\bibfield  {title} {\bibinfo {title} {Cylindrical {{Josephson}} tunneling},\ }\href {https://doi.org/10.1103/PhysRevB.19.1463} {\bibfield  {journal} {\bibinfo  {journal} {Phys. Rev. B}\ }\textbf {\bibinfo {volume} {19}},\ \bibinfo {pages} {1463} (\bibinfo {year} {1979})}\BibitemShut {NoStop}%
\bibitem [{\citenamefont {Bhushan}\ and\ \citenamefont {Sherrill}(1981)}]{Bhushan:PB81}%
  \BibitemOpen
  \bibfield  {author} {\bibinfo {author} {\bibfnamefont {M.}~\bibnamefont {Bhushan}}\ and\ \bibinfo {author} {\bibfnamefont {M.~D.}\ \bibnamefont {Sherrill}},\ }\bibfield  {title} {\bibinfo {title} {Cylindrical {{Josephson}} junctions},\ }\href {https://doi.org/10.1016/0378-4363(81)90670-7} {\bibfield  {journal} {\bibinfo  {journal} {Physica B+C}\ }\textbf {\bibinfo {volume} {107}},\ \bibinfo {pages} {735} (\bibinfo {year} {1981})}\BibitemShut {NoStop}%
\bibitem [{\citenamefont {Sherrill}(1981)}]{Sherrill:PLA81}%
  \BibitemOpen
  \bibfield  {author} {\bibinfo {author} {\bibfnamefont {M.~D.}\ \bibnamefont {Sherrill}},\ }\bibfield  {title} {\bibinfo {title} {Critical magnetic field of cylindrical {{Josephson}} junctions},\ }\href {https://doi.org/10.1016/0375-9601(81)90118-3} {\bibfield  {journal} {\bibinfo  {journal} {Physics Letters A}\ }\textbf {\bibinfo {volume} {82}},\ \bibinfo {pages} {191} (\bibinfo {year} {1981})}\BibitemShut {NoStop}%
\bibitem [{\citenamefont {Burt}\ and\ \citenamefont {Sherrill}(1981)}]{Burt:PLA81}%
  \BibitemOpen
  \bibfield  {author} {\bibinfo {author} {\bibfnamefont {P.~B.}\ \bibnamefont {Burt}}\ and\ \bibinfo {author} {\bibfnamefont {M.~D.}\ \bibnamefont {Sherrill}},\ }\bibfield  {title} {\bibinfo {title} {The dc {{Josephson}} current in cylindrical junctions},\ }\href {https://doi.org/10.1016/0375-9601(81)90232-2} {\bibfield  {journal} {\bibinfo  {journal} {Physics Letters A}\ }\textbf {\bibinfo {volume} {85}},\ \bibinfo {pages} {97} (\bibinfo {year} {1981})}\BibitemShut {NoStop}%
\bibitem [{\citenamefont {Wang}\ and\ \citenamefont {Xu}(1991)}]{Wang:JLTP91}%
  \BibitemOpen
  \bibfield  {author} {\bibinfo {author} {\bibfnamefont {S.~H.}\ \bibnamefont {Wang}}\ and\ \bibinfo {author} {\bibfnamefont {L.~D.}\ \bibnamefont {Xu}},\ }\bibfield  {title} {\bibinfo {title} {Behavior of dual superconducting cylinders in a magnetic field},\ }\href {https://doi.org/10.1007/BF00681528} {\bibfield  {journal} {\bibinfo  {journal} {J Low Temp Phys}\ }\textbf {\bibinfo {volume} {82}},\ \bibinfo {pages} {217} (\bibinfo {year} {1991})}\BibitemShut {NoStop}%
\bibitem [{\citenamefont {Hadfield}\ \emph {et~al.}(2003)\citenamefont {Hadfield}, \citenamefont {Burnell}, \citenamefont {Kang}, \citenamefont {Bell},\ and\ \citenamefont {Blamire}}]{Hadfield:PRB03}%
  \BibitemOpen
  \bibfield  {author} {\bibinfo {author} {\bibfnamefont {R.~H.}\ \bibnamefont {Hadfield}}, \bibinfo {author} {\bibfnamefont {G.}~\bibnamefont {Burnell}}, \bibinfo {author} {\bibfnamefont {D.-J.}\ \bibnamefont {Kang}}, \bibinfo {author} {\bibfnamefont {C.}~\bibnamefont {Bell}},\ and\ \bibinfo {author} {\bibfnamefont {M.~G.}\ \bibnamefont {Blamire}},\ }\bibfield  {title} {\bibinfo {title} {Corbino geometry {{Josephson}} junction},\ }\href {https://doi.org/10.1103/PhysRevB.67.144513} {\bibfield  {journal} {\bibinfo  {journal} {Phys. Rev. B}\ }\textbf {\bibinfo {volume} {67}},\ \bibinfo {pages} {144513} (\bibinfo {year} {2003})}\BibitemShut {NoStop}%
\bibitem [{\citenamefont {Clem}(2010)}]{Clem:PRB10}%
  \BibitemOpen
  \bibfield  {author} {\bibinfo {author} {\bibfnamefont {J.~R.}\ \bibnamefont {Clem}},\ }\bibfield  {title} {\bibinfo {title} {Corbino-geometry {{Josephson}} weak links in thin superconducting films},\ }\href {https://doi.org/10.1103/PhysRevB.82.174515} {\bibfield  {journal} {\bibinfo  {journal} {Phys. Rev. B}\ }\textbf {\bibinfo {volume} {82}},\ \bibinfo {pages} {174515} (\bibinfo {year} {2010})}\BibitemShut {NoStop}%
\bibitem [{\citenamefont {Zhang}\ \emph {et~al.}(2022)\citenamefont {Zhang}, \citenamefont {Lyu}, \citenamefont {Wang}, \citenamefont {Zhuo}, \citenamefont {Sun}, \citenamefont {Li}, \citenamefont {Shen}, \citenamefont {Liu}, \citenamefont {Qu},\ and\ \citenamefont {L{\"u}}}]{Zhang:CPB22}%
  \BibitemOpen
  \bibfield  {author} {\bibinfo {author} {\bibfnamefont {Y.}~\bibnamefont {Zhang}}, \bibinfo {author} {\bibfnamefont {Z.}~\bibnamefont {Lyu}}, \bibinfo {author} {\bibfnamefont {X.}~\bibnamefont {Wang}}, \bibinfo {author} {\bibfnamefont {E.}~\bibnamefont {Zhuo}}, \bibinfo {author} {\bibfnamefont {X.}~\bibnamefont {Sun}}, \bibinfo {author} {\bibfnamefont {B.}~\bibnamefont {Li}}, \bibinfo {author} {\bibfnamefont {J.}~\bibnamefont {Shen}}, \bibinfo {author} {\bibfnamefont {G.}~\bibnamefont {Liu}}, \bibinfo {author} {\bibfnamefont {F.}~\bibnamefont {Qu}},\ and\ \bibinfo {author} {\bibfnamefont {L.}~\bibnamefont {L{\"u}}},\ }\bibfield  {title} {\bibinfo {title} {Ac {{Josephson}} effect in {{Corbino-geometry Josephson}} junctions constructed on {{Bi2Te3}} surface},\ }\href {https://doi.org/10.1088/1674-1056/ac89d4} {\bibfield  {journal} {\bibinfo  {journal} {Chinese Phys. B}\ }\textbf {\bibinfo {volume} {31}},\ \bibinfo {pages} {107402} (\bibinfo {year} {2022})}\BibitemShut {NoStop}%
\bibitem [{\citenamefont {Beenakker}(1992)}]{Beenakker:92}%
  \BibitemOpen
  \bibfield  {author} {\bibinfo {author} {\bibfnamefont {C.~W.~J.}\ \bibnamefont {Beenakker}},\ }\bibfield  {title} {\bibinfo {title} {Three "universal" mesoscopic {{Josephson}} effects}\ }(\bibinfo {year} {1992})\ pp.\ \bibinfo {pages} {235--253},\ \Eprint {https://arxiv.org/abs/cond-mat/0406127} {arXiv:cond-mat/0406127} \BibitemShut {NoStop}%
\bibitem [{\citenamefont {Sabonis}\ \emph {et~al.}(2020)\citenamefont {Sabonis}, \citenamefont {Erlandsson}, \citenamefont {Kringh{\o}j}, \citenamefont {{van Heck}}, \citenamefont {Larsen}, \citenamefont {Petkovic}, \citenamefont {Krogstrup}, \citenamefont {Petersson},\ and\ \citenamefont {Marcus}}]{Sabonis:PRL20}%
  \BibitemOpen
  \bibfield  {author} {\bibinfo {author} {\bibfnamefont {D.}~\bibnamefont {Sabonis}}, \bibinfo {author} {\bibfnamefont {O.}~\bibnamefont {Erlandsson}}, \bibinfo {author} {\bibfnamefont {A.}~\bibnamefont {Kringh{\o}j}}, \bibinfo {author} {\bibfnamefont {B.}~\bibnamefont {{van Heck}}}, \bibinfo {author} {\bibfnamefont {T.~W.}\ \bibnamefont {Larsen}}, \bibinfo {author} {\bibfnamefont {I.}~\bibnamefont {Petkovic}}, \bibinfo {author} {\bibfnamefont {P.}~\bibnamefont {Krogstrup}}, \bibinfo {author} {\bibfnamefont {K.~D.}\ \bibnamefont {Petersson}},\ and\ \bibinfo {author} {\bibfnamefont {C.~M.}\ \bibnamefont {Marcus}},\ }\bibfield  {title} {\bibinfo {title} {Destructive {{Little-Parks Effect}} in a {{Full-Shell Nanowire-Based Transmon}}},\ }\href {https://doi.org/10.1103/PhysRevLett.125.156804} {\bibfield  {journal} {\bibinfo  {journal} {Phys. Rev. Lett.}\ }\textbf {\bibinfo {volume} {125}},\ \bibinfo {pages} {156804} (\bibinfo {year} {2020})}\BibitemShut {NoStop}%
\bibitem [{\citenamefont {Beenakker}\ and\ \citenamefont {{van Houten}}(1991)}]{Beenakker:PRL91}%
  \BibitemOpen
  \bibfield  {author} {\bibinfo {author} {\bibfnamefont {C.~W.~J.}\ \bibnamefont {Beenakker}}\ and\ \bibinfo {author} {\bibfnamefont {H.}~\bibnamefont {{van Houten}}},\ }\bibfield  {title} {\bibinfo {title} {Josephson current through a superconducting quantum point contact shorter than the coherence length},\ }\href {https://doi.org/10.1103/PhysRevLett.66.3056} {\bibfield  {journal} {\bibinfo  {journal} {Phys. Rev. Lett.}\ }\textbf {\bibinfo {volume} {66}},\ \bibinfo {pages} {3056} (\bibinfo {year} {1991})}\BibitemShut {NoStop}%
\bibitem [{\citenamefont {Beenakker}\ and\ \citenamefont {{van Houten}}(1992)}]{Beenakker:NaMS92}%
  \BibitemOpen
  \bibfield  {author} {\bibinfo {author} {\bibfnamefont {C.~W.~J.}\ \bibnamefont {Beenakker}}\ and\ \bibinfo {author} {\bibfnamefont {H.}~\bibnamefont {{van Houten}}},\ }\bibfield  {title} {\bibinfo {title} {{{THE SUPERCONDUCTING QUANTUM POINT CONTACT1}}},\ }in\ \href {https://doi.org/10.1016/B978-0-12-409660-8.50051-1} {\emph {\bibinfo {booktitle} {Nanostructures and {{Mesoscopic Systems}}}}},\ \bibinfo {editor} {edited by\ \bibinfo {editor} {\bibfnamefont {W.~P.}\ \bibnamefont {Kirk}}\ and\ \bibinfo {editor} {\bibfnamefont {M.~A.}\ \bibnamefont {Reed}}}\ (\bibinfo  {publisher} {Academic Press},\ \bibinfo {address} {Cambridge},\ \bibinfo {year} {1992})\ pp.\ \bibinfo {pages} {481--497}\BibitemShut {NoStop}%
\bibitem [{\citenamefont {Setiawan}\ and\ \citenamefont {Hofmann}(2022)}]{Setiawan:PRR22}%
  \BibitemOpen
  \bibfield  {author} {\bibinfo {author} {\bibfnamefont {F.}~\bibnamefont {Setiawan}}\ and\ \bibinfo {author} {\bibfnamefont {J.}~\bibnamefont {Hofmann}},\ }\bibfield  {title} {\bibinfo {title} {Analytic approach to transport in superconducting junctions with arbitrary carrier density},\ }\href {https://doi.org/10.1103/PhysRevResearch.4.043087} {\bibfield  {journal} {\bibinfo  {journal} {Phys. Rev. Res.}\ }\textbf {\bibinfo {volume} {4}},\ \bibinfo {pages} {043087} (\bibinfo {year} {2022})}\BibitemShut {NoStop}%
\bibitem [{\citenamefont {Kruti}\ and\ \citenamefont {Riwar}(2024)}]{Kruti:PRB24}%
  \BibitemOpen
  \bibfield  {author} {\bibinfo {author} {\bibfnamefont {D.}~\bibnamefont {Kruti}}\ and\ \bibinfo {author} {\bibfnamefont {R.-P.}\ \bibnamefont {Riwar}},\ }\bibfield  {title} {\bibinfo {title} {Interplay between evanescent scattering modes and finite dispersion in superconducting junctions},\ }\href {https://doi.org/10.1103/PhysRevB.110.224513} {\bibfield  {journal} {\bibinfo  {journal} {Phys. Rev. B}\ }\textbf {\bibinfo {volume} {110}},\ \bibinfo {pages} {224513} (\bibinfo {year} {2024})}\BibitemShut {NoStop}%
\bibitem [{\citenamefont {Schwiete}\ and\ \citenamefont {Oreg}(2010)}]{Schwiete:PRB10}%
  \BibitemOpen
  \bibfield  {author} {\bibinfo {author} {\bibfnamefont {G.}~\bibnamefont {Schwiete}}\ and\ \bibinfo {author} {\bibfnamefont {Y.}~\bibnamefont {Oreg}},\ }\bibfield  {title} {\bibinfo {title} {Fluctuation persistent current in small superconducting rings},\ }\href {https://doi.org/10.1103/PhysRevB.82.214514} {\bibfield  {journal} {\bibinfo  {journal} {Phys. Rev. B}\ }\textbf {\bibinfo {volume} {82}},\ \bibinfo {pages} {214514} (\bibinfo {year} {2010})}\BibitemShut {NoStop}%
\bibitem [{\citenamefont {Nyg{\aa}rd}()}]{Nygard:}%
  \BibitemOpen
  \bibfield  {author} {\bibinfo {author} {\bibfnamefont {J.}~\bibnamefont {Nyg{\aa}rd}},\ }\href@noop {} {\bibinfo {title} {Private {{Communication}}}}\BibitemShut {NoStop}%
\bibitem [{\citenamefont {{Everschor-Sitte}}\ \emph {et~al.}(2014)\citenamefont {{Everschor-Sitte}}, \citenamefont {Sitte},\ and\ \citenamefont {MacDonald}}]{Everschor-Sitte:JoAP14}%
  \BibitemOpen
  \bibfield  {author} {\bibinfo {author} {\bibfnamefont {K.}~\bibnamefont {{Everschor-Sitte}}}, \bibinfo {author} {\bibfnamefont {M.}~\bibnamefont {Sitte}},\ and\ \bibinfo {author} {\bibfnamefont {A.~H.}\ \bibnamefont {MacDonald}},\ }\bibfield  {title} {\bibinfo {title} {Half-metallic magnetism and the search for better spin valves},\ }\href {https://doi.org/10.1063/1.4893969} {\bibfield  {journal} {\bibinfo  {journal} {Journal of Applied Physics}\ }\textbf {\bibinfo {volume} {116}},\ \bibinfo {pages} {083906} (\bibinfo {year} {2014})}\BibitemShut {NoStop}%
\bibitem [{\citenamefont {{San-Jose}}(2025)}]{San-Jose:25a}%
  \BibitemOpen
  \bibfield  {author} {\bibinfo {author} {\bibfnamefont {P.}~\bibnamefont {{San-Jose}}},\ }\href {https://doi.org/10.5281/zenodo.4762963} {\bibinfo {title} {Pablosanjose/{{Quantica}}.jl}} (\bibinfo {year} {2025})\BibitemShut {NoStop}%
\bibitem [{\citenamefont {Pay{\'a}}(2025{\natexlab{a}})}]{Paya:25a}%
  \BibitemOpen
  \bibfield  {author} {\bibinfo {author} {\bibfnamefont {C.}~\bibnamefont {Pay{\'a}}},\ }\href {https://doi.org/10.5281/zenodo.15013533} {\bibinfo {title} {{{CarlosP24}}/{{FullShell}}.jl: {{Full-shell}} code updated for {{Josephson Junctions}}}},\ \bibinfo {howpublished} {Zenodo} (\bibinfo {year} {2025}{\natexlab{a}})\BibitemShut {NoStop}%
\bibitem [{\citenamefont {Pay{\'a}}(2025{\natexlab{b}})}]{Paya:25h}%
  \BibitemOpen
  \bibfield  {author} {\bibinfo {author} {\bibfnamefont {C.}~\bibnamefont {Pay{\'a}}},\ }\href {https://doi.org/10.5281/zenodo.15013478} {\bibinfo {title} {{{CarlosP24}}/{{SP43}}\_{{SNS}}\_{{Paper}}: {{Final Publication Release}}: {{Josephson Effects}} in {{Full-Shell Nanowires}}}},\ \bibinfo {howpublished} {Zenodo} (\bibinfo {year} {2025}{\natexlab{b}})\BibitemShut {NoStop}%
\bibitem [{\citenamefont {Danisch}\ and\ \citenamefont {Krumbiegel}(2021)}]{Danisch:JOSS21}%
  \BibitemOpen
  \bibfield  {author} {\bibinfo {author} {\bibfnamefont {S.}~\bibnamefont {Danisch}}\ and\ \bibinfo {author} {\bibfnamefont {J.}~\bibnamefont {Krumbiegel}},\ }\bibfield  {title} {\bibinfo {title} {Makie.jl: {{Flexible}} high-performance data visualization for {{Julia}}},\ }\href {https://doi.org/10.21105/joss.03349} {\bibfield  {journal} {\bibinfo  {journal} {Journal of Open Source Software}\ }\textbf {\bibinfo {volume} {6}},\ \bibinfo {pages} {3349} (\bibinfo {year} {2021})}\BibitemShut {NoStop}%
\bibitem [{\citenamefont {De~Gennes}(2018)}]{De-Gennes:18}%
  \BibitemOpen
  \bibfield  {author} {\bibinfo {author} {\bibfnamefont {P.-G.}\ \bibnamefont {De~Gennes}},\ }\href@noop {} {\emph {\bibinfo {title} {Superconductivity of Metals and Alloys}}}\ (\bibinfo  {publisher} {CRC Press},\ \bibinfo {address} {Boca Raton, FL},\ \bibinfo {year} {2018})\BibitemShut {NoStop}%
\end{thebibliography}%

\end{document}